\documentclass[12pt]{article}

\usepackage{dutchcal}

\usepackage[makeroom]{cancel}

\usepackage{latexsym}

\usepackage{longtable}

\usepackage{epsfig}

\usepackage[utf8]{inputenc}
\usepackage[english]{babel}
\usepackage{mathrsfs}

\usepackage{graphicx,bbm,psfrag}

\usepackage{amssymb, amsmath, amsthm,booktabs,mathtools}

\usepackage{bm}

\usepackage{color}
\usepackage{hyperref}
\hypersetup{
    colorlinks,
    citecolor=red,
    linkcolor=blue,
}

\usepackage[toc,page]{appendix}

\newcommand{\commutator}[2]{\left[#1,#2\right]}

\newcommand{\anticommuator}[2]{\left\{#1,#2\right\}}

\newcommand{\vev}[1]{\langle #1 \rangle}

\DeclareMathOperator{\pf}{\mathrm{Pf}}

\newcommand{\ud}          {\mathrm d}
\newcommand{\e}          {\mathrm e}

\newcommand\msf            {\mathsf}

\newcommand\ul[1]{\underline{#1}}

\newcommand{\bc}{\begin{center}}
\newcommand{\ec}{\end{center}}
\def\ba#1{\begin{array}{#1}\displaystyle}
\newcommand{\ea}{\end{array}}

\newcommand{\beq}{\begin{equation}}
\newcommand{\eeq}{\end{equation}}
\newcommand{\beqa}{\begin{eqnarray}}
\newcommand{\eeqa}{\end{eqnarray}}
\newcommand{\no}{\nonumber}
\newcommand{\n}{\nonumber\\}
\newcommand{\bi}{\begin{itemize}}
\newcommand{\ei}{\end{itemize}}

\def\lt#1{\left#1}
\def\rt#1{\right#1}

\def\b#1{\bar{#1}}
\def\frc#1#2{\frac{#1}{#2}}

\newcommand{\p}{\partial}

\newcommand{\bra}{\langle}
\newcommand{\ket}{\rangle}
\newcommand{\Z}{{\mathbb{Z}}}
\newcommand{\N}{{\mathbb{N}}}
\newcommand{\R}{{\mathbb{R}}}
\newcommand{\C}{{\mathbb{C}}}
\newcommand{\Tr}{{\rm Tr}}

\newcommand{\ep}{\epsilon}

\newcommand{\ri}{{\rm i}}
\newcommand{\re}{{\rm e}}
\newcommand{\dd}{{\rm d}}

\DeclareMathOperator{\arccosh}{arccosh}

\DeclareMathOperator{\imag}{Im}
\DeclareMathOperator{\real}{Re}

\newcommand{\rz}{\rm z}

\begin{document}

\begin{titlepage}

\begin{center}
{\large {\bf The hydrodynamic theory of dynamical correlation functions\\[0.1cm] in the XX chain}}

\vspace{1cm}

{ Giuseppe Del Vecchio Del Vecchio and Benjamin Doyon}
\vspace{0.2cm}

{\small\em
$^*$Department of Mathematics, King's College London, Strand, London WC2R 2LS, UK}

\end{center}

\vspace{1cm}
By the hydrodynamic linear response theory, dynamical correlation functions decay as power laws along certain velocities, determined by the flux Jacobian. Such correlations are obtained by hydrodynamic projections, and physically, they are due to propagating ``sound waves" or generalisation thereof, transporting conserved quantities between the observables. However, some observables do not emit sound waves, such as  order parameters associated to symmetry breaking. In these cases correlation functions decay exponentially everywhere, a behaviour not captured by the hydrodynamic linear response theory. Focussing on spin-spin correlation functions in the XX quantum chain, we first review how hydrodynamic linear response works, emphasising that the necessary fluid cell averaging washes out oscillatory effects. We then show how, beyond linear response, Euler hydrodynamics can still predict the exponential decay of correlation functions of order parameters. This is done by accounting for the large-scale fluctuations of domain walls, via the recently developed ballistic fluctuation theory. We use the framework of generalised hydrodynamics, which is particularly simple in this model due to its free fermion description. In particular, this reproduces, by elementary calculations, the exponential decay in the celebrated formulae by A.R. Its, A.G. Izergin, V.E. Korepin, N.A. Slavnov (1993) and by X. Jie (1998), which were originally obtained by intricate Fredholm determinant analysis; and gives a new formula in a parameter domain where no result was obtained before. We confirm the results by numerical simulations.



\vfill

\hfill \today

\end{titlepage}

\tableofcontents

\section{Introduction}

Understanding dynamical behaviours of many-body systems at large scales in space-time is a particularly difficult problem of many-body physics. These behaviours are controlled by emergent degrees of freedom and are often difficult to extract from microscopic calculations, even in models with strong mathematical structures such as integrable models.

For this purpose, hydrodynamics is an extremely powerful theory. For integrable systems, the focus here, Generalised Hydrodynamics (GHD) has provided a wealth of new results in recent years \cite{Castro_Alvaredo_2016,Bertini_2016}. According to the general principle of hydrodynamics, in stationary, homogeneous, clustering states -- the maximal entropy states, such as thermal or generalised Gibbs ensembles (GGE) -- the propagation of ``linear waves" gives simple predictions for the algebraic decay of dynamical two-point connected correlation functions \cite{denardis2021correlation}. Linear waves are small perturbations of the GGE in space-time, well described by linear response theory. They propagate according to Euler hydrodynamics, and mathematically lie in the tangent space to the state manifold (the tangent space is the space of extensive conserved quantities for this state). Dynamical correlation functions are then expected to decay exponentially in space-time, except at the available hydrodynamic velocities, elements of the spectrum of the flux Jacobian, where algebraic decay is found. The mechanism is that of the projection of the observables onto the conserved quantities carried by the linear waves, the so-called Boltzmann-Gibbs principle \cite{Spohn-book, demasi, kipnis}; a rigorous theory is given in \cite{doyon2020hydrodynamic} for quantum chains. In integrable systems, a continuum of velocities is available, and everywhere within this continuum, the decay is in time$^{-1}$; see the review \cite{denardis2021correlation}.

Hydrodynamic algebraic decay occurs whenever the observables ``couple" to at least one linear wave: this means that their expectation values vary along the state manifold, i.e. their susceptibilities are nonzero, see \cite{Doyon_2017,Doyon_2018}. However, certain observables do not vary in any direction on the manifold, and hence do not couple to linear waves. These may display exponential decay throughout space-time, and the Boltzmann-Gibbs principle is not useful. This may happen, for instance, for order parameters, or more generally twist fields (both highest-weight and descendants under the local observable algebra), in thermal states and other GGEs, as order is typically destroyed at nonzero entropy. Such fields -- which we will simply refer to as order parameters -- have zero expectation values (at least in a neighbourhood the GGE), hence zero susceptibility. There is no general hydrodynamic theory yet for describing the exponential decay of their correlation functions.

The goal of this paper is twofold: (1) we illustrate the hydrodynamic linear response theory for integrable systems, emphasising some of its subtleties and in particular the necessity for fluid-cell averaging; and (2) we propose and verify a new hydrodynamic theory for exponential decay of order parameters.

We use the example of the XX quantum chain:
\begin{equation}
    H=-\sum_{x\in\Lambda}\left[\sigma^1_{x}\sigma^1_{x+1}+\sigma^2_{x}\sigma^2_{x+1}-h\sigma^3_x\right]\label{eq:H_XX}
\end{equation}
where $\sigma^{1,2,3}_x$ are the Pauli matrices at site $x\in\Z$. We obtain results at infinite volumes, $\Lambda = \Z$. Hydrodynamic linear response is applied to correlation functions of $\sigma^3_x$ (``longitudinal"), while the new theory is applied to correlation functions of order parameters, which can be taken as $\sigma^{1}_x$ or $\sigma^{2}_x$ (``transverse"). This model is simple enough, having a free-fermionic description, so that the hydrodynamic principles can be verified against exact known asymptotic results and numerical analysis. Yet it is nontrivial enough to illustrate the important aspects. Like hydrodynamic linear response, the new theory we propose, being based on hydrodynamics, is fully generalisable to truly interacting models, integrable or not.

The new theory for order-parameter exponential decay is based on the ballistic fluctuation theory  developed in \cite{doyon2019fluctuations}, and the associated theory of twist-field correlation functions proposed there. This framework is a generalisation of the ideas introduced in the context of large deviations in conformal field theory \cite{Bernard_2016}. The physical principle is relatively simple: the exponential decay of the $\bra\sigma^+_x(t)\sigma^-_0(0)\ket$ correlation function is controlled, in part, by the number of domain walls crossing between the two observables, and their fluctuations, as first proposed for the Ising model in \cite{Sachdev_1996, Sachdev_1997, Buragohain_1999}. The large-deviation theory is applied to this extensive, fluctuating variable (the number of domain walls).  Because of the free-fermion structure underlying the XX chain, its Euler hydrodynamics, a special case of generalised hydrodynamics, is particularly simple, and hence the large-deviation theory takes a simple form. In fact, the specific form of the observables $\sigma^\pm_x$ means that they not only count domain walls, but also act as source and sink of domain walls. By the Jordan-Wigner transformation, they are fermionic descendants of the $U(1)$ twist field of the XX model. We explain how to correctly account for this, and how it gives, for certain values of parameters, an additional contribution to the exponential decay.

More recently, a number of studies appeared on the asymptotics of dynamical correlation functions in spin chains: in the transverse field Ising model for example a partial resummation of the form factor expansion leads to the asymptotic behaviour of the order parameter in the low-density regime \cite{essler2020}. Importantly, Fredholm determinant techniques continue to provide non-trivial results on large-time behaviour \cite{zhuravlev2021large, chernowitz2021dynamics, 10.21468/SciPostPhys.10.3.070} of correlation function in spin chains: this provides a bridge between hydrodynamic based techniques and asymptotic behaviour of a certain class of Fredholm determinant which might be useful even beyond the present context of applications. We leave more careful analysis and explicit connection for later studies.

\subsection{Main results}

{\em A. Longitudinal correlation and fluid-cell mean.} Hydrodynamic projections describe large space-time scales\footnote{Projections in fact occur on different Hilbert spaces depending on the wavelength and frequency \cite{doyon2020hydrodynamic,AmpelogiannisDoyon2021}; here we concentrate on zero frequency and infinite wavelength, and will discuss generalisation in a forthcoming work.}.  In general, it is not sufficient to take the limit of large space-time separation in the correlation: an appropriate averaging over a mesoscopic cell {\em in space-time} is required by the general theory \cite{doyon2020hydrodynamic}, in order to cancel potential oscillations. Depending on the model, the fluid-cell mean described in \cite{doyon2020hydrodynamic} may in fact be simplified. As there is no general theory yet for when and how this simplification occurs, it is interesting to determine precisely the form fluid-cell means may take in explicit examples.

In terms of the scaled space-time variables $\b x\in\R,\;\b t\in\R^+$, we find that a weak form of fluid-cell averaging is sufficient for the longitudinal correlator in the XX model. For all rays $\xi = \b x/\b t\in\R$, we consider the fluid-cell mean
\beq\label{averagetime}
	\overline{\sigma}_{\ell \b x}^{3}(\ell \b t)
	= \frac{1}{2\ell_0 } \int_{-\ell_0}^{\ell_0} \dd s\,\sigma_{\ell \b x}^{3}(\ell \b t + s)
\eeq
as well as, for comparison, a ``ray" mean, where the average is taken along the ray $\xi$ (similar to that used in \cite{10.21468/SciPostPhys.4.6.045}), see \eqref{average}. Here and below $\vec\sigma_x(t) = \e^{\ri H t}\vec \sigma_x \e^{-\ri Ht}$. Both means involve one-dimensional fluid cells. Hydrodynamic predictions occur when $\ell_0$ is mesoscopic and $\ell$ macroscopic: in the limit $\ell_0,\ell\to\infty$ with $0\ll\ell_0\ll\ell$.

We show that for all $\xi\in\R$,
\beq\label{corrzz}
	\bra\overline{\sigma}_{\ell \b x}^{3}(\ell \b t)\sigma_0^{3}(0)\ket^{\rm c} \sim
	\ell^{-1} S(\b x,\b t)\qquad (0\ll\ell_0\ll\ell)
\eeq
where $\vev{\cdots}$ is the expectation in an arbitrary GGE (see \eqref{eq:free_GGE}). We find a nonzero result in the time-like cone $|x|\leq 4t$:
\beq
	S(x,t) =\begin{cases} \displaystyle \frac{1}{2\pi }\frac{n_+(1-n_+) + n_-(1-n_-)}{\sqrt{t^2-(x/4)^2}}\,, &\xi=x/t \in(-4,4)
		\\\displaystyle
		0\,,  &\mbox{otherwise}
		\end{cases}
	\label{eq:corrzz_asymptotics}
\eeq
where
\beq
n_\pm = \frc1{e^{ w(k_\pm)}+1} \label{eq:generalized_filling}
\eeq
is the generalised Fermi factor. Here $k_\pm\equiv k_\pm(\xi)\in(-\pi,\pi)$ are the two distinct values of $k$ solving $v(k) =\xi$ with $v(k)=E'(k) = 4\sin(k)$ being the velocity of the quasi-particles, where $E(k)=2(h-2\cos(k))$ is the dispersion relation of the Hamiltonian \eqref{eq:H_XX}. The ordinary thermal case at inverse temperature $\beta$ is recovered using $w(k) = \beta E(k)$.

The result \eqref{corrzz} with \eqref{eq:corrzz_asymptotics} indicates that for $|\xi|<4$ a well-defined power-law asymptotic in $\ell^{-1}$ emerges, for $|\xi|>4$ the decay is faster than $\ell^{-1}$, and for $|\xi|=4$ it is slower. It is obtained by Wicks theorem, and, as we show, agrees with the predictions from hydrodynamics, confirming that a weak notion of fluid-cell mean is sufficient. It is important to remark that without fluid-cell averaging, the asymptotic \eqref{corrzz} is incorrect, as there are additional oscillatory terms, see \eqref{eq:corrzz_asymptotics_GGE}. In section \ref{secthydrocomparison}, we also show that the ray mean is a good fluid-cell mean except for an infinite set of rays of measure zero, where oscillatory terms remain; this is the case generically for one-dimensional means involving the discrete space direction. We finally remark that in fact, no averaging is required in the region $|x|\geq 4t$ for \eqref{corrzz} to hold, and that at $|\xi|=4$, the power law is $\ell^{-2/3}$ and it is non-oscillatory (see \eqref{xi4}).

{\em B. Exponential decay of order-parameter correlation.} The proposed hydrodynamic theory for order parameters allows us to obtain the exponential decay of the more technically involved ``transverse" correlation function. In this case, we obtain
\beq
	\lt|\bra\sigma_{\ell \b x}^+(\ell \b t)\sigma_0^-(0)\ket\rt|\asymp
	\e^{\ell F(\b x,\b t)}
	\label{eq:exp_decay_init}
\eeq
in all regions of space-time. The formula means that the leading large-$\ell$ asymptotic behaviour of the logarithm of both sides agree. Although, we obtain certain oscillating factors as well, taking the absolute value omits them as we want to focus on exponential decay only for the moment. Note that by the unbroken spin-flip symmetry,  it is not necessary to take the connected correlation function, and $\bra\sigma_{x}^+(t)\sigma_0^-(0)\ket^{\rm c} = \bra\sigma_{x}^+(t)\sigma_0^-(0)\ket = \bra\sigma_{x}^1(t)\sigma_0^1(0)\ket/2 = \bra\sigma_{x}^2(t)\sigma_0^2(0)\ket/2$.

For the correlation function \eqref{eq:exp_decay_init} in thermal states, known approaches in the XX model and other spin models with free-fermion structures are either approximate \cite{Sachdev_1996,Sachdev_1997,Buragohain_1999}, or exact but technically involved \cite{Colomo1993,G_hmann_2017,PhysRevB.100.155428}. The hydrodynamic theory we propose is much more direct, and translates the simple physical principles underlying exponential decay into an exact formula. Interestingly, to our knowledge, at space-like distances $|x|>4t$, exact results have been obtained in the literature only for moderate magnetic field $|h|<2$ (gapless regime). In the gapped case $|h|>2$, we find a new expression for $F(\b x,\b t)$. Overall, at inverse temperature $\beta$, and without loss of generality considering $t>0$,
\beqa\label{eq:therm_F}
F(x,t)=\lt\{
\begin{aligned}
	&f_{x,t}[\beta E] &&(|\xi|\leq 4)\\ 
	& |x|f_{1,0}[\beta E] &&(|\xi|> 4,\ |h|\leq 2)\\
	& -|x|\,{\rm min}\big(\arccosh(h/2),  M_\xi\big)+|x|f_{1,0}[\beta E]
	&&(|\xi|>4,\ |h|>2)
\end{aligned}
\rt.\quad .
\eeqa
Here, $M_\xi = \arccosh(\xi/4)-\sqrt{1-\frc{16}{|\xi|^2}}$ and we have defined a free-energy-type functional of the state, thermal in this particular case, as (see also eq.\ \eqref{Flambda} for a generalised version)
\begin{equation}
	f_{x,t}[\beta E] = \int_{-\pi}^\pi \frc{\dd k}{2\pi}\,
	\lt|x-v(k)t \rt|
	\log \Big|\tanh\frc{\beta E(k)}2\Big| .
\end{equation}
This agrees with the exponential part of the celebrated formulae by A.R. Its, A.G. Izergin, V.E. Korepin, N.A. Slavnov \cite{its1993temperature}, and by X. Jie \cite{Jie_Thesis}, in their stated parameter domains: $|h|<2$ in the whole spacetime for the former, $|h|>2$ and $|x|>4t$ for the latter.

In the work \cite{its1993temperature}, the correlation function is written as a product of two contributions: one coming from the Jordan-Wigner string, represented as a Fredholm determinant, and another contribution depending on a ``potential". Below we will make a similar ansatz that leads to formula \eqref{eq:therm_F}. We believe comparison between certain formulae of Ref. \cite{its1993temperature} and ours are illuminating. The representation of correlation functions as quantities satisfying non-linear integrable PDEs, and using Fredholm determinants, is exact. Then, on one side, the asymptotic of these can be obtained by application of non-linear steepest descent \cite{10.2307/2946540} to the associated Riemann-Hilbert problem \cite{korepin_bogoliubov_izergin_1993}, while on the other, our results show that there are purely Euler hydrodynamic arguments for such asymptotic. For convenience we collect some of the main observations from the PDE / Fredholm determinant side in Appendix \ref{comparison_its}.

Finally, note how the formula for $F(x,t)$ takes a different appearance in the space-like gapped regime hitherto not studied; this is, in fact, an analytic continuation in $h$ from the gapless regime $|h|<2$.The above formula has been verified numerically in the literature \cite{Derzhko_2000}, surprisingly, only in the regime $h<2$. We provide verifications in a larger regime of parameters and in both space-like and time-like regions. We also obtain formulae in arbitrary GGEs, see Section \ref{ssectexpo} and Eq.~\eqref{fullresulttransverse} in Appendix \ref{appsaddle}.

\subsection{Organisation of the paper} In section \ref{XXmodel} we review the basic solution to the XX model and state the problem. In section \ref{sectlongitu} we review the hydrodynamic linear response theory (hydrodynamic projection theory), and show that it correctly predicts the asymptotics of the longitudinal correlation function. In section \ref{secttransverse} we review the ballistic fluctuation theory, explain how it can be used to evaluate the transverse correlation functions in the XX model, and compare the results with numerical simulations. Finally, we conclude in section \ref{conclu}.

\section{The XX model and statement of the problem}\label{XXmodel}

The XX spin chain \eqref{eq:H_XX} is one the simplest examples of exactly solvable one-dimensional models, a special case of the more general XY spin model introduced and solved in \cite{LIEB1961407}. Consider the system on a finite periodic lattice $\Lambda = \{0,1,\ldots,N-1\}$. The Hamiltonian is diagonalisable by means of a Jordan-Wigner transformation  and Fourier transform. The relevant spin matrices are written as
\beq
	\sigma^+_x=\frac12(\sigma_x^1+\ri\sigma_x^2)
	=\exp\left(\ri\pi\sum_{y=0}^{x-1}a^\dagger_y a_y\right)a^\dagger_x,\quad\sigma^-_x = (\sigma^+_x)^\dag,\quad
	\sigma^3_x=2a^\dagger_x a_x- 1
	\label{eq:jw}
\eeq
in terms of canonical complex fermions $a_x$, with $\{a_x^\dag,a_y\}=\delta_{x,y},\;\{a_x,a_y\}=0$. This preserves the $su(2)$ algebra and structure of its spin-$1/2$ representation. The Hamiltonian is made quadratic, taking a different form on the sectors with even and odd fermion numbers,
\begin{equation}
    H =  P_+ H_+ + P_- H_-, \quad P_\pm = \frac{1}{2}\Big(1\pm(-1)^{\Omega}\Big)\label{eq:hamiltonian_splitting}
\end{equation}
where the projectors are expressed in terms of the fermion number $\Omega = \sum_{x=0}^{N-1} a^\dag_xa_x$. The even and odd Hamiltonians are
\begin{equation}
    H_\pm = -2\sum_{x=0}^{N-1}\left[a^\dagger_x a_{x+1}+a^\dagger_{x+1} a_x + h a^\dagger_x a_x\right] + h N
    \label{hamilfermions}
\end{equation}
where, implicitly, the boundary condition is anti-periodic (even sector, $+$) or periodic (odd sector, $-$), $a_{x+N} = (-1)^{\Omega+1}a_x$. A Fourier transform diagonalises the Hamiltonian,
\beq
	a_x = \frc{1}{\sqrt{N}}\sum_{k\in\Gamma_\pm}  \e^{\ri kx} c_k,\quad
	\{c_k^\dag,c_l\} = \delta_{k,l},\quad\{c_k,c_l\}=0
	 \label{eq:fourier_basis_discrete}
\eeq
where the momentum sets are such that the (anti-)periodicity condition is satisfied, $\Gamma_+ = \{j\Delta k:j=\lfloor -N/2\rfloor,\ldots,\lfloor N/2\rfloor-1\}$ and $\Gamma_- = \{j\Delta k:j=\lfloor -N/2\rfloor+1/2,\ldots,\lfloor N/2\rfloor-1/2\}$ for $\Delta k = \frac{2\pi}{N}$ (with $|\Gamma_\pm|=N$). This gives the dispersion relation $E(k)$:
\beq
	H_\pm = \sum_{k\in\Gamma_\pm}
	E(k) c_k^\dag c_k,\quad E(k) = 2(h-2\cos(k)).
	\label{eq:dispersion_relation}
\eeq
In each sector, the Fourier modes evolve as
\begin{equation}\label{sectorevol}
	c_k(t):= \re^{\ri H_\pm t} c_k\re^{-\ri H_\pm t} = \e^{-\ri E_kt}c_k\, ,\quad c_k^\dag(t):=\re^{\ri H_\pm t} c_k^\dag\re^{-\ri H_\pm t} = \e^{\ri E_k t}c^\dag_k\qquad (k\in\Gamma_\pm).
\end{equation}

The states of interest are the generalised Gibbs ensembles (by definition, these are the states that are stationary, homogeneous, and clustering at large distances), with the density matrix fully fixed by a function $W:\Gamma_\pm\ni k\to W_k\in\R$,
\beq\label{eq:GGE}
	\rho = \frc1Z\Big( P_+ \e^{-\sum_{k\in\Gamma_+} W_k c^\dag_k c_k}
	+ P_- \e^{-\sum_{k\in\Gamma_-} W_k c^\dag_k c_k}\Big)
\eeq
where $Z$ is such that $\Tr\rho=1$. The case
\beq\label{thermal}
	W_k = \beta E(k)
\eeq
is the thermal state at temperature $\beta^{-1}$.

We are interested in the infinite-length limit $N\to\infty$, where we have, in both sectors,
\beq
	a_x = \int_{-\pi}^\pi \frc{\dd k}{\sqrt{2\pi}}\,\e^{\ri kx} c(k),\quad
	\{c^\dag(k),c(l)\} = \delta(k-l),\quad\{c(k),c(l)\}=0,\quad
	x\in\Z.\label{eq:fourier_basis_continuum}
\eeq
Assuming that there is a function $w(k)$ that is continuous or continuous by part, and such that $w(k) = W_k\ \forall\,k\in\Gamma_\pm,\,\forall\,N$,  this simplifies the density matrix to
\beq
	\rho = \frc1Z \exp\left[-\int_{-\pi}^\pi \dd k\,w(k)c^\dag(k) c(k)\right].\label{eq:free_GGE}
\eeq
The correlation functions whose asymptotics we are looking to evaluate are
\beqa\label{longitudinal}
	\mbox{(longitudinal)}\qquad \bra \sigma_x^3(t)\sigma_0^3(0)\ket^{\rm c} &=& \Tr \big(\rho\, \sigma_x^3(t)\sigma_0^3(0)\big) - \Big(\Tr\big(\rho\, \sigma_0^3\big)\Big)^2\\ 
	\mbox{(transverse)}\qquad 
	\label{transverse}\bra \sigma_x^+(t)\sigma_0^-(0)\ket &=& \Tr \big(\rho\, \sigma_x^+(t)\sigma_0^-(0)\big)
\eeqa
where the nomenclature refers to the direction of the spin component with respect to the magnetic field. We are looking for the asymptotic regime
\beq\label{asymptotic}
	x=\lfloor \ell \b x\rfloor,\quad t = \ell \b t, \quad \ell\to\infty.
\eeq
More precisely, we wish to evaluate the asymptotic \eqref{corrzz} under the fluid-cell means \eqref{averagetime} and \eqref{average}, and the asymptotic \eqref{eq:exp_decay_init}.

Below, for simplicity we assume that $w(k)$ is a continuous and periodic function on the interval $k\in[-\pi,\pi]$ -- that is, continuous on the interval seen as a topological circle. In fact, all results hold as well if it is continuous by part on the circle, as long as its evaluation at a point of discontinuity $k_*$ is assigned to the average of the left- and right-limits $w(k_*) = \frc12 (\lim_{k\to k_*^-}w(k) + \lim_{k\to k_*^+}w(k))$, although this requires additional justifications that we will omit.

\section{Longitudinal correlators from hydrodynamic projections}\label{sectlongitu}

The leading asymptotics of the longitudinal correlation function $\bra \sigma_x^3(t)\sigma_0^3(0)\ket^{\rm c}$ in space-time can be predicted by the standard hydrodynamic linear response theory.  In this section we explain how this is done, and we compare with the result of the elementary computation using Wick's theorem, confirming the hydrodynamic theory and leading to  \eqref{eq:corrzz_asymptotics}. We show that the general form of the fluid-cell averaging, described in \eqref{meanobservable}, can here be simplified to \eqref{averagetime} and \eqref{average} for this hydrodynamic result to hold.

\subsection{Fluid-cell averages and hydrodynamic projections}\label{hydroeuler}

Correlation functions in many-body quantum and classical systems can be analysed at the Euler scale using the hydrodynamic description of the system. The Euler scale is that in which space and time are taken to be large, simultaneously. The hydrodynamic theory gives predictions for correlation functions of local (or quasi-local) observables at the Euler scale in stationary, homogeneous, clustering states like those described in \eqref{eq:free_GGE} for free theories. The latter are states which are invariant under time and space translations, and in which averages of observables factorise in the limit where they are placed at large spatial separation. Here we assume that clustering is fast enough, but we omit a detailed description of the requirements on the state; see for instance \cite{doyon2020hydrodynamic} for a rigorous treatment of Euler-scale correlation functions, \cite{perfetto2020eulerscale} for extension to inhomogeneous settings and \cite{denardis2021correlation} for a review of the theory of correlation functions in GHD, of which the discussion below is a special case.

The prediction from hydrodynamics is based on the available conservation laws admitted by the system. Consider a quantum chain on $\Z$ admitting $N$ conservation laws
\begin{equation}
	\p_t q_i(x,t) + j_i(x+1,t)-j_i(x,t) =  0, \label{eq:conservatopm_law_micro}
\end{equation}
where $q_i$ and $j_i$ are the conserved densities and currents, respectively. We assume for simplicity that the state is invariant under the action of all total charges $Q_i = \sum_{x\in\Z} \, q_i(x)$, that is $\bra[Q_i,\cdots]\ket = 0$, although this is not a necessary condition for the hydrodynamic theory to apply. Consider the following covariance matrices:
\beq
	\mathsf C_{ij} = \sum_{x\in\Z}\bra q_i(x) q_j(0)\ket^{\rm c},\quad
	\mathsf B_{ij} = \sum_{x\in\Z}\bra j_i(x) q_j(0)\ket^{\rm c}.
\eeq
Both matrices are symmetric; this is evident for $\mathsf C$, less so for $\mathsf B$ but a proof can be found in \cite{Doyon_2020} for instance. Consider also the ``flux Jacobian"
\beq
	\mathsf A_i^{~j} = \sum_k\mathsf B_{ik} \mathsf C^{kj}
\eeq
where we denote the inverse matrix with upper indices, $\sum_k\mathsf C^{ik}\mathsf C_{kj} = \delta^i_j$. For any local observable $o$, construct the vector
\beq\label{onepoint}
	\mathsf V^o_i = \sum_{x\in\Z}\bra q_i(x)o(0)\ket^{\rm c}.
\eeq
In particular, for conserved densities and currents we have $\mathsf V^{q_i}_k = \msf C_{ik}$ and $\mathsf V^{j_i}_k = \msf B_{ik}$.

The hydrodynamic prediction is for the correlation function of fluid-cell means $\overline{o}_{1,2}(x,t)$  of local observables $o_1(x,t)$ and $o_2(x,t)$. There are various ways of expressing the fluid-cell means (see \cite{doyon2020hydrodynamic}), but one way is to average over an appropriate space-time fluid cell. For our purpose, consider a family of closed time intervals $I_x\subset \R$ (which may be single points), one for each position $x\in\Z$ on the chain. Then a fluid cell is a set $\Upsilon = \cup_{x=-\ell_1}^{\ell_2} (x, I_x) \subset \Z\times \R$ in space-time, with $\ell_1,\ell_2\geq 0$, and the fluid-cell mean is the average over the fluid cell $\Upsilon+(x,t)$:
\beq\label{meanobservable}
	\overline{o}(x,t) = \frc1{\ell_2+\ell_1+1}
	\sum_{y=-\ell_1}^{\ell_2} \frc1{|I_y|}\int_{I_y}\dd s
	\,o(x+y,t+s).
\eeq
The fluid cell is taken to be ``mesoscopic". That is, the parameters $\ell_1,\,\ell_2$ and $I_x$ are taken to be dependent on an ``observation scale" $\ell$, such that the linear extent $\ell_0$ of the fluid cell, say $\ell_0 = {\rm max}\{\ell_1, \,\ell_2,\, |t|: t \in \cup_{x\in [-\ell_1,\ell_2]} I_x\}$, is monotonically increasing with $\ell$ but much smaller, $\lim_{\ell\to\infty} \ell_0/\ell =0$. The hydrodynamic prediction is that for $\ell_1,\,\ell_2$ and the intervals $I_x$ growing fast enough with $\ell$ (typically $\ell_0\rightarrow + \infty$ fast enough, but within the mesoscopic constraint), the correlation function of fluid-cell means, times the scale $\ell$ of the space-time positions, has a limit expressible solely in terms of the hydrodynamic matrices $\mathsf C,\, \mathsf A$ and the vectors $\mathsf V^{o_1},\,\mathsf V^{o_2}$, as follows:
\beq\label{corrfct}
	S_{o_1o_2}(\b x, \b t) := \lim_{\ell\to\infty} \ell \bra \overline{o}_1(\ell \b x,\ell \b t) o_2(0,0)\ket^{\rm c} = \mathsf V^{o_1} \cdot \mathsf C^{-1} \delta(\b x - \mathsf A \b t)\mathsf V^{o_2}.
\eeq
Note that it is sufficient, by space-time translation invariance, to average over the positions of a single observable. Expression \eqref{corrfct} is a hydrodynamic projection formula, where $\mathsf V^{o_1}$ and $\mathsf V^{o_2}$ represent the projection of $o_1$ and $o_2$ onto conserved quantities, and $ \mathsf C^{-1}\delta(\b x - \mathsf A \b t)$ represents the propagation of hydrodynamic modes. This precise statement is expected to hold but has not be shown rigorously. A mathematically rigorous expression of a slightly different but related hydrodynamic projection statement is proven, see \cite{doyon2020hydrodynamic}.

The flux Jacobian can be shown to be diagonalisable and to have a real spectrum. For $N$ finite, the limit in \eqref{corrfct} is a generalised function of the scaled space-time coordinates $\b x,\,\b t$, concentrated on the space-time rays of velocities equal to the eigenvalues $v_i^{\rm eff}$ of $\mathsf A$. These eigenvalues are therefore interpreted as the velocities of propagation of linear disturbances on top of the fluid, which give rise to the leading (Euler-scale) correlations. If the state takes the Gibbs form, with density matrix
\beq\label{gge}
	\rho=\frc1{Z}\exp\left( -\sum_i\beta^iQ_i\right),
\eeq
then, by using the chain rule to change to the $\beta^i$ coordinates, one observes that $\msf{A}_i^{~j} = \frac{\p\bra j_i\ket}{\p \bra q_j\ket}$, which justifies the name ``flux Jacobian".

The application of these general principles to GHD was done in \cite{Doyon_2017}, see the review \cite{denardis2021correlation} and references therein. As the XX model has a free fermion structure, GHD simplifies drastically, see \cite{PhysRevB.96.220302}. The state \eqref{eq:free_GGE} is stationary and homogeneous, and one can show that it is clustering as long as $w(k)$ has appropriate analytic properties; in particular, if $w(k)$ is analytic on the real line, then the two-point functions of fermions are exponentially decaying. The states \eqref{eq:free_GGE} are in fact the generalised Gibbs ensembles, and on these states one may apply the hydrodynamic theory of correlation functions.

Given the state \eqref{eq:free_GGE}, one may construct the occupation function
\beq\label{eq:occupation_GGE}
	n(k) = \frc1{1+e^{w(k)}}.
\eeq
All local conserved quantities, such as the Hamiltonian, have the form
\beq\label{conscharges}
	Q_i = \int_{-\pi}^\pi \dd k\,h_i(k) c^\dag(k)c(k)
\eeq
where the function $h_i(k)$ is the one-particle eigenvalue of $Q_i$. The index $i$ indexes any chosen (discrete, say) basis of the set of conserved quantities. The average densities may be evaluated by going back to a finite system, with discrete values of $k$, density matrix \eqref{eq:free_GGE} and $Q_i = \sum_{k\in\Gamma_+\cup\Gamma_-} h_i(k) c_k^\dag c_k$, and by evaluating $\lim_{N\to\infty} \bra Q_i\ket/N$. The currents also take a universal form \cite{PhysRevB.96.220302,FagottiLocally}. The result is the standard one of GHD (see e.g. \cite{Castro_Alvaredo_2016, Bertini_2016} as well as \cite{Borsi_2021, PhysRevLett.125.070602, PhysRevX.10.011054, spohn2020collision, VY18, yoshimura2020collision} for other derivations),
\beq
	\bra q_i\ket =\int_{-\pi}^\pi \frc{\dd k}{2\pi}\,n(k)h_i(k),\quad
	\bra j_i\ket =\int_{-\pi}^\pi \frc{\dd k}{2\pi}\,v(k)n(k)h_i(k)
\eeq
where the group velocity is
\beq\label{velocity}
	v(k) = E'(k) = 4\sin(k).
\eeq
The covariance matrices are evaluated by standard statistical mechanics methods, for instance
\beq
	\mathsf C_{ij} = \int_{-\pi}^\pi \frc{\dd k}{2\pi}\,n(k)(1-n(k))h_i(k)h_j(k).
\eeq
The flux Jacobian is diagonalised by passing to the  continuous basis of the Fourier modes, and one has
\beq
	\sum_j \mathsf A_i^{~j}h_j(k) = v(k)h_i(k).
\eeq
Thus the spectrum of $\mathsf A$ is the interval $[-4,4]$. 

The hydrodynamic result for correlation functions of conserved densities can then be written. Using the continuous diagonal basis, one can simplify the expression \eqref{corrfct} into a single integral:
\beqa
	S_{q_i,q_j}(x, t) &=& \int_{-\pi}^\pi \frc{\dd k}{2\pi}\,\delta(x-v(k)t) n(k)(1-n(k))h_i(k)h_j(k)\\
	&=& \lt\{\ba{ll} 
	\displaystyle
	\frc1{2\pi \sqrt{16 t^2-x^2}}\sum_{\sin (k_\pm) = \xi/4\atop
	k_\pm\in[-\pi,\pi)}  n(k_\pm)(1-n(k_\pm)) h_i(k_\pm)h_j(k_\pm) & (|x/t|\leq 4) \n
	0 & (|x/t|>4).\ea\rt.
	\label{corrfcthydro}
\eeqa
For $|x/t|< 4$, as the result is a finite ordinary function, and not a generalised function, the hydrodynamic prediction is therefore that the mean-observable correlation function decays as $1/\ell$,
\beq\label{qqhydro}
	\bra \overline{q}_i(\ell \b x,\ell \b t) q_j(0,0)\ket\sim \ell^{-1} S_{q_i,q_j}(\b x,\b t) \qquad (\ell\to\infty,\ |x/t|<4).
\eeq
For $|x/t|>4$, the result for $S_{q_i,q_j}(x, t)$ vanishes, meaning that the decay is faster than $\ell^{-1}$. For $|x/t|=4$, the result diverges, meaning that the decay is slower than $\ell^{-1}$.

\subsection{Hydrodynamics reproduces the Wick-theorem, saddle-point result}\label{secthydrocomparison}

As the total spin shifted by a constant is twice the total number of fermions,
\beq\label{Q0}
	Q_0:=\frc12 \sum_{x=0}^{N-1} (\sigma_x^3+{\bf 1}) = 
	\Omega,\qquad
	q_0(x) \equiv q(x) = \frc12 (\sigma^3_x+1) = a^\dag_x a_x
\eeq
and thus a conserved quantity, we may use the above theory in order to predict the asymptotics of the correlation function of its density $\bra q(x,t)q(0,0)\ket$ at the Euler scale. The constant shift does not affect connected correlation functions, which then boils down to $ \frc14 \bra \sigma_x^3(t)\sigma_0^3(0)\ket^{\rm c}$. In the general result \eqref{corrfcthydro}, the required ingredients, besides the occupation function, is the one-particle eigenvalue $h_0(k)\equiv h(k)$ corresponding to this conserved quantity. Clearly, for a single fermion the eigenvalue of $\Omega$ is $h(k)=1$, and hence the prediction reproduces \eqref{eq:corrzz_asymptotics}.

The longitudinal correlation function, as given in \eqref{longitudinal}, can also be evaluated by a direct microscopic calculation using Wick's theorem. The last formula in \eqref{eq:jw} and space-time translation invariance give the connected longitudinal correlation function
\begin{equation}
	\vev{\sigma^3_{x}(t)\sigma^3_{0}(0)}^{\rm c}
	 = 4\vev{a^\dag_x(t) a_x(t) a^\dag_0(0) a_0(0)}-4\vev{a^\dag_0(0) a_0(0)}^2
\end{equation}
which gives by Wick's theorem
\beq
	\vev{\sigma^3_{x}(t)\sigma^3_{0}(0)}^{\rm c}
	 = 4\vev{a^\dag_x(t)  a_0(0) }\vev{ a_x(t)a^\dag_0(0)}.
\eeq
We work directly in the thermodynamic limit as prescribed in Eq.~\eqref{eq:fourier_basis_continuum}. Using $\vev{c^\dag(k) c(l)} = \delta(k-l)n(k)$ and $\vev{c(k) c^\dag(l)} = \delta(k-l)(1-n(k))$, we obtain
\beq\label{res}
	\vev{\sigma^3_{x}(t)\sigma^3_{0}(0)}^{\rm c}
	= 4 \int_{-\pi}^{\pi} \int_{-\pi}^{\pi} \frc{\dd k \dd l}{(2\pi)^2} \e^{-\ri (k-l)x +\ri (E(k) - E(l))t }n(k)(1-n(l)).
\eeq
We consider $t>0$ for simplicity.

Inside the light-cone, that is when $\xi=\b x/\b t \in(-4,4)$, the asymptotic regime \eqref{asymptotic} is obtained by a stationary phase analysis taking $x$ and $t$ both of order $\ell\gg 1$. The result, derived below, is a power law decay with power $-1$, in agreement with \eqref{corrzz}. Away from the light-cone, $|\xi|>4$, as the stationary phases do not lie on the integration region, the vanishing is faster, thus the second line of  \eqref{eq:corrzz_asymptotics} holds. Its precise form depends on the analytic structure of $n(k)$. For instance, if $n(k)$ is analytic on $[-\pi,\pi]$, the asymptotic is obtained by contour deformation and the vanishing is at least exponential; this is because everywhere on $[-\pi,\pi]$, the phase derivative is either positive or negative (since it is never zero), and hence there is a purely imaginary direction in which an infinitesimal displacement gives a real decaying exponential. We omit the general analysis of the case $|\xi|>4$ here for simplicity. At $|\xi|=4$, the stationary phase analysis must be modified, leading to a slower decay, as we explain below.

\subsubsection{Case $|\xi|<4$} In the asymptotic regime \eqref{asymptotic}, no matter the specific fluid-cell mean, we may perform a saddle point analysis. The exponential in \eqref{res} admits, for both the $k$ and $l$ integration, a set of two stationary points $k_\pm=k_\pm(\xi)$ on the integration intervals, given by $k_+(\xi)={\rm arcsin}(\xi/4)\in(-\pi/2,\pi/2)$ and $k_- = {\rm sgn}(k_+)\pi-k_+$, which solve the saddle-point equation $E'(k_\pm) = \xi$. Using the fact that $E''(k_\pm) = \pm \sqrt{16-\xi^2}$ and $E(k_+)-E(k_-) = -2\sqrt{16- \xi^2}$, the result may be written, with $x=\xi t$, as
\beqa	\label{eq:corrzz_asymptotics_GGE}
	\lefteqn{\vev{\sigma^3_{ x}(t+s)\sigma^3_{0}(0)}^{\rm c}} \\ && =  \frac{2}{ \pi \sqrt{16 t^2-x^2}}\sum_{a=\pm}
	n_a\Big(1-n_a
+a \ri \,(1-n_{-a}) (-1)^x \e^{-2a\ri(k_+ x + (t+s)\sqrt{16 -\xi^2})}+ O\left(t^{-1}\right)\Big)\no
\eeqa
where $n_\pm = n(k_\pm)$.

We show that after fluid-cell averaging this reproduces \eqref{eq:corrzz_asymptotics}. We consider two types of fluid-cell averaging. We show that (I) the ``time" mean \eqref{averagetime} is a valid mean for all rays $\xi\in(-4,4)$, and that (II) 
the ``ray" mean
\beq\label{average}
	\overline{\sigma}_{\ell \b x}^{3}(\ell \b t)
	=
	\frac{1}{2\ell_0 } \sum_{y= -\ell_0+1}^{\ell_0} \sigma_{\ell \b x + y}^{3}(\ell \b t + y /\xi)
\eeq
is a valid mean for all rays in a certain set, $\xi\in(-4,4)\setminus\Xi$ with
\beq\label{setXi}
	\Xi = \{0\}\cup\{\xi_*\in(-4,4)\setminus\{ 0\}:2\,{\rm arcsin}(\xi_*/4) + 2\sqrt{16 \xi_*^{-2}-1} \in 2\pi \Z + \pi\}
\eeq
(where ${\rm arcsin}(\xi_*/4)\in (-\pi/2,\pi/2)$). This is the interval $(-4,4)$, but excluding a countable set, of measure zero.

In the case (I), the fluid-cell mean \eqref{averagetime} is a specialisation of the general form \eqref{meanobservable}, with $\ell_1=\ell_2=0$ and $I_0 = [-\ell_0,\ell_0]$. In the case (II), it is a specialisation to the choice $\ell_1 = -\ell_0+1$, $\ell_2 = \ell_0$ and the minimal choice of one-point time intervals $I_x = \{x/\xi\}$. The value of $\ell_0$ does not have to be taken ``large enough", and any $0\ll \ell_0\ll \ell$ will work. Both are therefore minimal expressions of the fluid-cell mean, in some way with a fluid cell that is the ``least extended possible". For $\xi\in\Xi$, in the case (II) this minimal mean {\em does not give the hydrodynamic prediction}, but with an additional minimal averaging over rays, the prediction is reproduced (again this can be recast into a specialisation of the general fluid-cell mean \eqref{meanobservable}).

Case (I). In \eqref{eq:corrzz_asymptotics_GGE} we separate the $\ell$-scaling variables $x=\xi t = \ell\b x$, which stay on the ray $\xi$, from the $\ell_0$-scaling variable $s$, the addition to the time variable, which goes away from it, and use $\ell$ as the large parameter for the stationary point analysis. We choose a ray $\xi\in(-4,4)$ and set $x= \ell \b x,\,t= \ell \b t$, and preform the average over $s$. The non-oscillating term is independent of $s$ and gives
\beq
	\frc2{\pi\sqrt{16 t^2-x^2}}\sum_{a=\pm} n_a(1-n_a) + O(\ell^{-1}).
\eeq
In order to evaluate the oscillating term, we perform the fluid cell averaging, which gives the vanishing
\beq
	\lim_{\ell_0\to\infty} \frc1{2\ell_0} \int_{-\ell_0}^{\ell_0} \dd s\,\exp\big(
	-a\ri(2k_+ x + 2(t+s)\sqrt{16 -\xi^2})\big) = 0
\eeq
uniformly on $x,t$, using the fact that $|\xi|<4$. This correctly reproduces \eqref{corrzz} with \eqref{eq:corrzz_asymptotics}, in accordance with the hydrodynamic prediction.

Case (II). Everywhere within the fluid cell, both $x$ and $t$ are large and stay on the ray $\xi$, with $x = \xi t  = \ell \b x (1 + O(\ell_0/\ell))$. In this case, we may do the stationary phase analysis uniformly everywhere within the cell, the scale $O(\ell) = O(x) = O(t)$ being the large parameter. We choose a ray $\xi\in(-4,4)\setminus \{0\}$, set $x= \ell \b x + y,\,t= \ell \b t + y/\xi$, and consider the fluid-cell averaging \eqref{average} with $\ell_0\ll \ell$. In this limit, the non-oscillating terms in \eqref{eq:corrzz_asymptotics_GGE} may be evaluated by using
\begin{align*}
	\frac{1}{2\ell_0 } \sum_{y= -\ell_0+1}^{\ell_0} \frac{1}{\ell \bar{t} + y/\xi}\frac{1}{\sqrt{16-\xi^2}}=\frac{1}{\ell}\frac{1}{\sqrt{16\b t^2-\b x^2}}\left(1+O(\ell_0/\ell)\right).
\end{align*}
This immediately reproduces \eqref{eq:corrzz_asymptotics}, and we must show that the fluid-cell averaging vanishes on the oscillating terms. In order to do so, consider
\beq
(-1)^{\ell \b x + y}\exp\big(-a\ri (2k_+ + 2\sqrt{16 \xi^{-2}-1})(\ell \b x + y)\big).
\eeq
The fluid-cell sum over $y$ vanishes if, and only if, 
\beq
	\xi\not\in\Xi
\eeq
where $\Xi$ is given in \eqref{setXi}: if $\xi\in\Xi$, the terms are in fact not oscillating for $y\in\Z$, and add up to a finite contribution. There are infinitely many rays $\xi\in(-4,0)$, and $\xi\in(0,4)$, which break the condition. For instance, as $\xi$ increases from $-4$ to $0$, $k_+$ increases from $-\pi/2$ to 0, while $\sqrt{16 \xi^{-2}-1}$ increases from $0$ to $\infty$; therefore, all values $2\pi n + \pi$ for $n=1,2,\ldots$ will be crossed. However, as this is a set of isolated values $\xi_*$ of $\xi$, of measure zero, under additional averaging
\beq
	\frc1{2\ep} \int_{\xi_*-\ep}^{\xi_*+\ep} \dd \xi\label{eq:ray_average}
\eeq
the oscillatory terms contribution vanishes. As the sum over $y$ can be uniformly bounded for all rays, it can be evaluated on the integrand in \eqref{eq:ray_average} by the dominated convergence theorem, giving a result that is zero except for a set of measure zero (which is the single-point set $\{\xi_*\}$ as $\ep\to0$).

\subsubsection{Case $|\xi|=4$}  In this case the precise asymptotics is not predicted by hydrodynamics, but the Wick-theorem result allows us to evaluate it. We look for the leading asymptotics in the region $x=\ell \b x,\,t=\ell \b t$, $\ell\to\infty$. In this case $E''(k_\pm)=0$ and therefore in the exponent in \eqref{res} we must expand to 3rd order. Using  $k_\pm = k_+ = {\rm sgn}(\xi) \,\pi/2$ (there is a single stationary point), $E'''(k_+) = -4\,{\rm sgn}(\xi)$ and $x = 4\,{\rm sgn}(\xi) t$, we evaluate
\beqa
	\lefteqn{\int_{-\pi}^\pi \frc{\dd k }{2\pi} \e^{- \ri k x + \ri E(k) t } f(k)} && \n
	&&\sim \e^{-2 \ri \pi t + \ri E(\pi/2) t} \int_{-\pi}^\pi \frc{\dd k }{2\pi} \e^{ - 4 \ri\,{\rm sgn}(\xi) \,t k^3/3! } f(k_++k) \n
	&&\sim t^{-1/3} \e^{-2 \ri \pi t + \ri E(\pi/2) t} \int_{0}^\infty \frc{\dd k }{2\pi} \e^{- k^3 } \big(1/u + 1/u^*\big)f(k_+)\n
	&&\sim t^{-1/3} \e^{-2 \ri \pi t + \ri E(\pi/2) t}\frc{\Gamma(4/3)}{2\pi} 2\,{\rm Re}(1/u)f(k_+)
\eeqa
where $u = (4 \ri\,{\rm sgn}(\xi)/3!)^{1/3}$, and the integral contour has been deformed into a wedge in the complex plane around $k=0$ in order to have a sum of two convergent real integrals. The result is therefore
\beq\label{xi4}
	\vev{\sigma^3_{x}(t)\sigma^3_{0}(0)}^{\rm c}\sim
	t^{-2/3}\Big(\frc{\Gamma(4/3)}{\pi}3^{5/6}2^{-4/3}\Big)^2 n_+(1-n_+)\qquad (x = \pm 4t).
\eeq
This agrees with the hydrodynamic prediction \eqref{qqhydro}: the decay $t^{-2/3}$ is slower than $t^{-1}$.

\section{Transverse correlators from ballistic fluctuations}\label{secttransverse}

The prediction from the hydrodynamic linear response theory for the transverse correlation function $\bra \sigma_x^+(t)\sigma_0^-(0)\ket$, see section \ref{hydroeuler}, is a vanishing Euler-scale asymptotics. Indeed, according to formula \eqref{corrfct}, the correlation function is proportional to the integrated correlator $\mathsf V_i^{\sigma^\pm} = \sum_{x\in\Z}\bra q_i(x)\sigma^\pm_0\ket^{\rm c}$, Eq.~\eqref{onepoint}, involving the operator $\sigma_0^\pm$ at position 0, and the conserved densities $q_i(x)$. But all conserved charges \eqref{conscharges} preserve the total $\rz$-component of the spin, as does the GGE \eqref{eq:free_GGE}, and thus $\mathsf V^{\sigma^\pm}_i=0$; formula \eqref{corrfct} gives zero. This means that the decay of the fluid-cell average of the correlator must be faster than $1/t$.

In fact, it is known \cite{its1993temperature,Jie_Thesis} that the decay is exponential along any ray in space-time. Such exponential decays are not predicted by current hydrodynamic theories. However, recently \cite{doyon2019fluctuations} it was proposed that the leading exponent for the decay of correlation functions of certain types of observables, referred to as twist fields, may be predicted by Euler hydrodynamics, via the {\em ballistic fluctuation theory} (BFT). By the Jordan-Wigner transformation \eqref{eq:jw}, the transverse spin observables $\sigma_x^\pm$ are simply related to such twist fields. In this section, we explain how combining this general theory with simple Wick-theorem calculations, one can reproduce the known exponential decay of $\bra \sigma_x^+(t)\sigma_0^-(0)\ket$, and its generalisation to arbitrary parameter range and GGE. We verify our predictions by numerical calculations, and further explore the twist fields themselves, showing agreement with the BFT. 

\subsection{Ballistic fluctuation theory}\label{subsec:ballistic_fluctuation_theory}

\begin{figure}[ht]
	\begin{center}
		\includegraphics[width=0.8\textwidth]{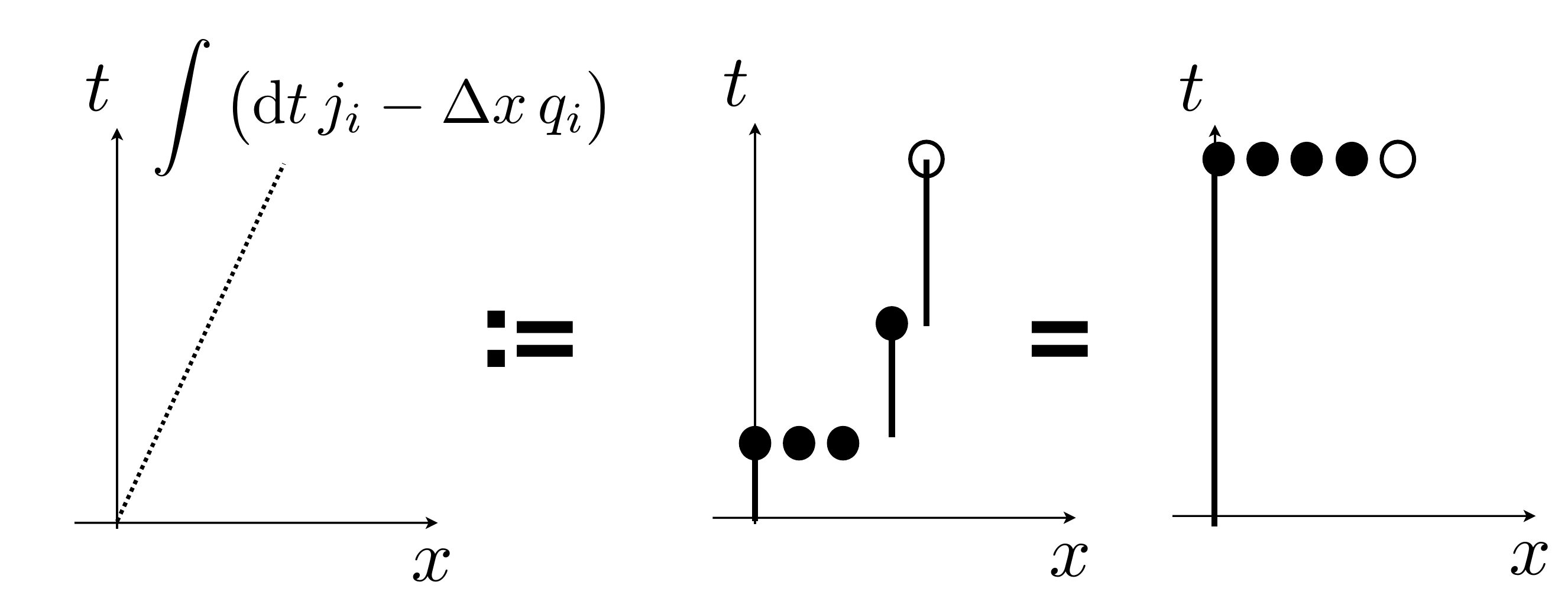}
	\end{center}
	\caption{The integral of the two-current on a discrete space. It is defined by integrating over times and summing over positions. The result is independent of the path chosen. In the middle, the case shown is an integral from $(0,0)$ to $(4,t)$ with $t_0=0,\,t_1=t_2=t_3>0, t=t_5>t_4>t_3$, see \eqref{eq:pathindependence}.}
	\label{fig:pathindependence}
\end{figure}

We now explain the basics of the ballistic fluctuation theory \cite{doyon2019fluctuations}. Consider the general hydrodynamic setup of section \ref{hydroeuler}. The BFT is concerned with the evaluation of the average
\beq\label{gdef}
g(\lambda,\ell;\b x,\b t) = \big\bra \exp \big(\lambda \Omega_{i_*}(\lfloor \ell\b x\rfloor,\ell \b t)\big)\big\ket,\quad
\Omega_{i_*}(\lfloor \ell\b x\rfloor,\ell \b t) = \int_0^{\ell\b t} \dd s\, j_{i_*}(0 ,s) - \sum_{y=0}^{\lfloor \ell \b x\rfloor-1} q_{i_*}(y ,\ell\b t)
\eeq
for $\lambda\in\C$, in the limit $\ell\to\infty$.  Note that the right hand side of the second equation in \eqref{gdef} can be interpreted as an integral over a path in spacetime $\mathbb{Z}\times \mathbb{R}$ where the integration measure is the standard product measure on  $\Z\times\R$. The integrand is the perpendicular-component of the 2-current $\dd t \,j_{i_*} - \Delta x\, q_{i_*}$ along the path (with $\Delta x=1$), see Fig. \ref{fig:pathindependence}. Here $q_{i_*}$ refers to a particular conserved density of interest associated with a conserved current $j_{i_*}$, both fixed once and for all so we suppress the index in the notation for the generating function. The integral is explicitly taken over the ``upper-left corner" path in space-time, but it is path-invariant by the conservation law \eqref{eq:conservatopm_law_micro}, in the sense that
\beq\label{eq:pathindependence}
\int_{(0,0)}^{(x,t)}\big(\dd t \,j_{i_*} - \Delta x\, q_{i_*}\big) := \sum_{n=0}^{x-1}\big( \int_{t_n}^{t_{n+1}} \dd s\,j_{i_*}(n,s) - q_{i_*}(n,t_{n+1})\big)
\eeq
is independent from the choice of $0=t_0\leq t_1\leq\ldots\leq t_x=t$ for a given end-point $(x,t)\in\Z\times\R$, see Fig.~\ref{fig:pathindependence}. Here $x = \lfloor \ell \b x\rfloor,\,t=\ell \b t$. The quantity $\Omega_{i_*}(x,t)$ is the total 2-current of the charge $Q_{i_*}$ evaluated along the ray $(0,0)\to(x,t)$. The state is taken to be some maximal entropy state of the Gibbs form \eqref{gge}, with Lagrange parameters $\beta^i$. Then, $g(\lambda,\ell;\b x,\b t)$ is the generating function for the cumulants of this total 2-current.

We are interested in the asymptotic regime \eqref{asymptotic}, and in particular in the exponential behaviour (renaming $\b x,\,\b t\to x,t$)
\beq\label{gasymp}
	g(\lambda,\ell; x,t) \asymp \re^{\ell f_{\lambda; x, t}[\boldsymbol \beta]}\qquad(\ell\to\infty).
\eeq
The function in the exponent may be complex, and thus this includes oscillatory terms. On the right-hand side, we explicitly write the dependence on the state specified by the Lagrange parameters $\boldsymbol \beta$. The precise limit to be evaluated is
\beq
	f_{\lambda; x, t}[\boldsymbol \beta] = \lim_{\ell\to\infty}\ell^{-1}
	\log g(\lambda,\ell; x,t).
\eeq

The BFT predicts that 
\begin{equation}
	f_{\lambda; x, t}[\boldsymbol \beta]=\int_0^\lambda\dd \lambda'\big(
	 t \,\mathsf{j}_{i_*}(\lambda';\xi)- x\,\mathsf{q}_{i_*}(\lambda';\xi)
	\big)
	\label{eq:SCGF2}
\end{equation}
where $\mathsf j_{i_*}(\lambda;\xi)$ and $\mathsf q_{i_*}(\lambda;\xi)$ are (G)GE averages of the current and density evaluated in a $\lambda$-dependent state described by $\beta^j(\lambda;\xi)$, which depends on the ray $\xi = x/t$ and obey the BFT flow equation
\beq\label{flow}
	\p_\lambda \beta^j(\lambda;\xi) = {\rm sgn}\big(x\,{\bf 1}-t\,\mathsf A(\lambda;\xi)\big)_{i_*}^{~j},\qquad
	\beta^j(0;\xi) = \beta^j,\quad \xi = x/t.
\eeq
The sign of the matrix is understood by diagonalisation as usual, and makes sense as $\mathsf A$ has real spectrum. The flow equation is derived under certain assumptions; it is expected to hold if the spectrum of $\mathsf A$ does not contain $x/t$, and also for (a large class of) integrable models, which possess a continuous spectrum.

The function $f_{\lambda; x, t}[\boldsymbol \beta]$ has an interpretation in terms of large-deviation theory. In the classical context, for instance at $x=0$ (where it is sufficient to consider $t=1$), it encodes the large deviations of the total current $J_{i_*}^{(\ell)} = \int_0^{\ell} \dd s\,j_{i_*}(0,s)$ in the time interval $[0,\ell]$: it is the ``full counting statistics", or scaled cumulant generating function, for the total amount of charge $Q_{i_*}$ that has crossed the point $x=0$, between times 0 and $\ell$. By large-deviation principle, the probability that $J_{i_*}^{(\ell)}$ takes the large value $\ell \mathcal j$ is generically exponentially decaying with $\ell$ as
\beq
	\mathbb{P}(J_{i_*}^{(\ell)}=\ell\mathcal j)\asymp \e^{-\ell I(\mathcal j)}
\eeq
for some large-deviation function $I(\mathcal j)$ (with $I(\mathsf j_{i_*})=0$ and $I(\mathcal j)>0 \,\,\,\mbox{if}\,\,\, \mathcal j\neq\mathsf j_{i_*}$, where $\mathsf j_{i_*} = \bra j_{i_*}\ket$). The function $f_{\lambda; 0, 1}[\boldsymbol \beta]$, as a function of $\lambda$, is the Legendre-Fenchel transform of $I(\mathcal j)$. The BFT gives a nontrivial prediction for this Legendre-Fenchel transform.

Likewise, at $x=1$ and $t=0$, a similar analysis applies where $\Delta Q_{i_*}(t)$ is replaced by the random variable $Q_{i_*}|_0^{\lfloor \ell\rfloor} = \sum_{s=0}^{\lfloor \ell\rfloor} q_{i_*}(s,0)$, the total charge on the spatial interval $[0,\ell]$. In this case, the BFT formulae above imply that $f_{\lambda; 1, 0}[\boldsymbol \beta]$ is the difference of specific free energies $\mathsf f[\boldsymbol \beta]$ at different states as follows:
\beq
	f_{\lambda; 1, 0}[\boldsymbol \beta] =\lim_{\ell\to\infty} \ell^{-1} \log \Big\bra \exp -\lambda\int_0^\ell \dd s\,  q_{i_*}(s,0)\big)\Big\ket =  \mathsf f[\boldsymbol \beta + \lambda \boldsymbol \delta_{i_*}]
	- \mathsf f[\boldsymbol \beta]
\eeq
where $(\boldsymbol \delta_{j})^i = \delta^i_{j}$.
As a consequence, the corresponding large-deviation function $I(\mathcal q)$ is simply related to the thermodynamic entropy density as a function of the charge $\mathcal q$.

The BFT thus gives predictions for the full counting statistics of total transported charges at large times, and total charges on large intervals, purely in terms of hydrodynamic and thermodynamic quantities, in any maximal entropy state\footnote{The BFT is based on an assumed fast enough decay of correlation functions at large distances, and thus it applies in finite-entropy states; ground states may need further analysis.}.

In integrable models, using GHD, the above can be translated into expressions in terms of the thermodynamic Bethe ansatz, see \cite{Myers_2020}. For the XX model, this simplifies thanks to the free fermion structure. With the GHD description as given in section \ref{hydroeuler}, and further using the free-energy function $\mathsf F(k) = -\log(1+e^{-w(k)})$, the results of the BFT specialise as follows. The GGE along the flow is described by the function
\beq\label{floww}
	w(\lambda;\xi;k) = w(k)+\lambda\, {\rm sgn}\big(x- v(k)t\big)\,h_{i_*}(k)
\eeq
where $v(k) = 4\sin (k)$ is the group velocity \eqref{velocity}, and the scaled cumulant generating function is
\beq\label{Flambda}
	f_{\lambda; x, t}[w] = \int_{-\pi}^\pi \frc{\dd k}{2\pi}\,
	|x-t\, v(k) |
	\log \Big(\frc{1+e^{-w(\lambda;\xi;k)}}{1+e^{-w(k)}}\Big)
\eeq
where we use $w$ instead of the Lagrange parameters $\boldsymbol \beta$ to specify the state.

\subsection{Transverse correlators and spin fluctuations}\label{section:transverse correlation}

The basis for the hydrodynamic theory of the correlation function $\bra \sigma^+_x(t)\sigma^-_0(0)\ket$ is the realisation that it is related to large deviations of the fermion number 2-current along the ray $(0,0)\to(x,t)$.

The first step is the extension of the usual Jordan-Wigner (JW) strings to ``space-time JW strings". By the JW transformation \eqref{eq:jw}, 
\beq\label{JWcorr0}
	\bra \sigma^+_x(0)\sigma^-_0(0)\ket 
	= \big\bra a_x^\dag \re^{\ri\pi \Omega_0(x,0)}\, a_0\big\ket
\eeq
as per the definition in \eqref{gdef}. Here we identify the total fermion number with the $i=0$ charge \eqref{Q0}, and use
\beq
	\sum_{y=0}^{x-1} a^\dag_y a_y = -\Omega_0(x,0)
\eeq
(the sign is unimportant in the exponential in \eqref{JWcorr0}, as the fermion number is an integer). It turns out that a similar formula holds for the large-scale exponential decay of the time-dependent correlator, 
\beq\label{corrOmega}
	\bra \sigma^+_x(t)\sigma^-_0(0)\ket 
	\asymp \big\bra a_x^\dag(t) \re^{\ri\pi \Omega_0(x,t)}\, a_0(0)\big\ket\qquad (x=\lfloor\ell \b x\rfloor,\ t=\ell \b t,\ \ell\to\infty),
\eeq
where, on the right-hand side, $\Omega_0(x,t)$ is defined in \eqref{gdef}, and the time evolution in $a_x^\dag(t)$ is under the sector-specific free-fermion Hamiltonian, as per \eqref{sectorevol}; as the limit of infinite chain length has been taken, the choice of sector does not matter. Formula \eqref{corrOmega} is derived below. We also establish that exact, instead of asymptotic, equality holds in \eqref{corrOmega} not only at $t=0$, but also at $x=0$. In fact, the following expression is exact for all $x,t$'s:
\begin{equation}\label{corrOmegaproduct}
	\sigma_x^+(t) \sigma_0(0)=  a_x^\dag(t)
	\exp\left(\ri\pi\sum_{y=0}^{x-1}q(y,t)\right) 
	\exp \lt(\ri \pi \int_0^t \dd s \,j(0,s) \rt) a_0(0)
\end{equation}
where we recall that $q(x,t)$ is the fermion density \eqref{Q0}, and $j(x,t)=j_0(x,t)$ is its associated current as per \eqref{eq:conservatopm_law_micro}.

We consider the correlator $\bra \sigma^+_x(t)\sigma^-_0(0)\ket$ and wish to establish formula \eqref{corrOmega}. For simplicity, we concentrate on the expectation being taken in the odd-fermion sector, projected onto by $P_-$; the argument is similar in the other sector. Then, according to the JW transformation \eqref{eq:jw}, in $\bra \sigma^+_x(t)\sigma^-_0(0)\ket$ we may set $\sigma_0^-(0) = a_0$ and
\beq\label{sigmap}
\sigma_x^+(t) = \re^{\ri H_- t}a^\dagger_x\exp\left(\ri\pi\sum_{y=0}^{x-1}a^\dagger_y a_y\right) \re^{-\ri H_+ t} = a_x^\dag(t)
\exp\left(\ri\pi\sum_{y=0}^{x-1}q(y,t)\right) 
\re^{\ri H_- t}\re^{-\ri H_+ t}
\eeq
where we use $q(x) = a^\dag _x a_x$, and where the fermion time-evolution is with respect to $H_-$ (see \eqref{sectorevol}).

We now show that, in the limit of an infinite chain $N\to\infty$, we have
\beq\label{technical}
\re^{\ri H_- t} \re ^{-\ri H_+t} =
\exp \lt(\ri \pi \int_0^t \dd s \,j(0,s) \rt).
\eeq
For this purpose, let us denote $Q^+ = \lim_{z\to\infty} Q|_0^z$ with $Q|_0^z = \sum_{x=0}^{z-1} q_0(x)$, and let denote by  $H_\pm|_y^z$ the fermionic hamiltonians $H_\pm$ where in \eqref{hamilfermions} the sum is for $x$ from $y$ to $z-1$. Then, one can check that for every $y\in\N$ and  $z\gg y$,
\beq
\re^{-\ri H_+|_{-y}^y t} = \re^{\ri \pi Q|_0^z} \re^{-\ri H_-|_{-y}^y t}\re^{-\ri\pi Q|_0^z}.
\eeq
Therefore,
\beq\label{technicaly}
\re^{\ri H_-|_{-y}^y t} \re ^{-\ri H_+|_{-y}^yt} =
\re^{\ri \pi Q|_0^z(t)} \re^{-\ri \pi Q|_0^z}
\eeq
where $Q|_0^z(t)$ is evolved with respect to $H_-|_{-y}^y$. Note that $q(x)$ is still a conserved density for time evolution by $H_-|_{-y}^y$. Its associated current is the operator
\beq\label{j0}
j(x) = 2\ri (a^\dag_x a_{x-1} - a^\dag_{x-1}a_x)
\eeq
for time evolution by $H_-$ for all $|x|\ll y$, but it is identically 0 for all $|x|\gg y$, because at these positions the evolution by $H_-|_{-y}^y$ is trivial. Therefore, using the conservation law \eqref{eq:conservatopm_law_micro} (for $H_-|_{-y}^y$), we have
\beq\label{q0current}
Q|_0^z(t)= Q|_0^z + \int_0^t \dd s \,j(0,s).
\eeq
The limit $y\to\infty$ on the left-hand side of \eqref{technicaly} exists, as $H_-|_{-y}^y - H_+|_{-y}^y$ is supported on the sites $-1,0$, and thus by the Baker-Campbell-Hausdorff formula, the left-hand side becomes an operator supported with exponential accuracy around the position $0$. Similarly, the limit $z\gg y\to\infty$ on the right-hand side of \eqref{technicaly}  also exists, using \eqref{q0current} and the fact that, by the Lieb-Robinson bound \cite{LiebRobinson}, the time-evolved current is supported on a finite number of site (at most proportional to $t$) with exponential accuracy. In this limit, $j(0,s)$ is now evolved with respect to $H_-$. We obtain
\begin{equation}\label{technical2}
	\re^{\ri H_- t} \re ^{-\ri H_+t} =
	\exp \lt(\ri \pi \int_0^t \dd s \,j(0,s) + \ri \pi  Q^+\rt) \exp\lt(-\ri \pi Q^+\rt).
\end{equation}
By contour deformation, we may rewrite
\beq\label{rewrite}
\int_0^t \dd s \,j(0,s) = \int_0^t \dd s \,j(z,s)
+ Q|_0^z(t) - Q|_0^z.
\eeq
By the Lieb-Robinson bound, the operator $I = \int_0^t \dd s \,j(z,s)$  is supported, with exponential accuracy, on $[z-vt,z+vt]$, where $v$ is the Lieb-Robinson velocity. Therefore, the projection of $I$ on negative positions is exponentially small as $z$ is made large. As both $H_-$ and $j(z)$ preserve fermion number, the operator $I$ is also fermion-number-preserving. Therefore, for $z$ large enough, all operators on the right-hand side of \eqref{rewrite} are fermion-number-preserving and supported on non-negative positions. Hence they commute with $Q^+$ (with exponential accuracy). The Baker-Campbell-Hausdorff formula applied to \eqref{technical2} then gives \eqref{technical}, which is exact as the result is independent of $z$.
Using \eqref{technical} in expression \eqref{sigmap}, and similar arguments on the even-fermion sector (projector $P_+$) we obtain \eqref{corrOmegaproduct}.
We emphasise that this is an exact formula, and we expect that it be possible to make it rigorous with currently known theorems in the context of the algebraic formulation of quantum chains.

We now argue that the correct leading asymptotic behaviour at large $x$ is \eqref{corrOmega}. The argument is based on the idea that the commutator of local observables at large distances in space-time tends to zero. This follows from the Lieb-Robinson bound if we understand local observables as those supported on finite number of sites, and if we take large space-like distances (again as determined by the Lieb-Robinson velocity). However, for our argument, we need to extend the concept of locality to ``semi-local" observables, end-points of appropriate infinite strings (such as in the Jordan-Wigner transformation, or more generally twist fields), and to assume that vanishing also holds uniformly enough in time-like regions as well. We note that in time-like regions, the weaker statement of ``almost-everywhere ergodicity" has been shown rigorously in thermal states for all quantum spin chains with short-range interactions \cite{doyon2020hydrodynamic,ampelogiannis2021almost}, based on exponential decay of correlations in space \cite{Araki}.

The observables
\begin{equation}
	Q^+(x) := \sum_{y=x}^\infty q(y)
\end{equation}
are fermion-number-preserving, and commute with any fermion-number-preserving local operator supported at positions $z$ far enough from $x$ (this fact we have already used in our derivation above). Therefore, $Q^+(x)$ may be adjoined to the space of fermion-number-preserving local operators\footnote{It is shown in \cite{doyon2020hydrodynamic} that such observables are elements of the Gelfand-Naimark-Segal Hilbert space, and that they have good locality properties; this gives a partial justification of the present loose argument.}. This space also includes $j(x)$. Commutators of operators within this space vanish at large space-like separations. Assuming that vanishing also holds in time-like directions\footnote{The weaker statement of vanishing of correlations under long-time averaging, valid almost everywhere in space-time including time-like directions, is shown in \cite{AmpelogiannisDoyon2021}, which serves to partially justify our present hypothesis.}, it is then a simple matter to see that, by the Baker-Campbell-Hausdorff formula,
\beqa
\lefteqn{\exp\left(\ri\pi\sum_{y=0}^{x-1}q(y,t)\right) 
	\exp \lt(\ri \pi \int_0^t \dd s \,j(0,s) \rt)} && \n 
&=&
\exp\left(\ri\pi Q^+(x,t) - \ri \pi Q^+(0,t) \right) 
\exp \lt(\ri \pi \int_0^t \dd s \,j(0,s) \rt) \n
&\asymp&
\exp\left(\ri\pi Q^+(x,t) - \ri \pi Q^+(0,t) +
\ri \pi \int_0^t \dd s \,j(0,s) \rt);
\eeqa
that is, the neglected commutators do not contribute to the leading long-time terms in the exponential. This gives \eqref{corrOmega}.

\subsection{Factorisation}
We see in \eqref{corrOmega} that the transverse correlation function is related to the two-point function of fermion operators, modified by a counting function for the integrated fermion number 2-current: the number of fermions lying, and the current flowing, between them. This counting function, the space-time JW string $\re^{\ri\pi \Omega_0(x,t)}$, gives a factor $-1$ for every fermion counted or flowing by. As the fermion number is the number of spins in the direction up, the transverse correlator is a modification of a local fermion correlator by the fluctuation of the total spin and its current between them.

We note that a formula similar to \eqref{corrOmega} is well-known for the classical, two-dimensional Ising model, a statistical model simply related to this XX quantum chain but in Euclidean time \cite{ZuberItzykson,SchroerTruong}. However, we believe this is the first time the formula is established in real time for the quantum model. This is a crucial formula for our argument on the exponential decay of $\bra \sigma^+_x(t)\sigma^-_0(0)\ket$ in space-time.

The next step is to use the BFT, which gives the asymptotic \eqref{gasymp} for the expectation of the space-time JW string, along with simple Wick-theorem arguments, in order to obtain from \eqref{corrOmega} the leading exponential decay. For this purpose, we use the BFT concept of a flow on states, with the flow equation \eqref{flow}. We propose the asymptotic factorisation property
\beq\label{corraa}
	\big\bra a_x^\dag(t) \re^{\lambda\Omega_0(x,t)}\, a_0(0)\big\ket
	\asymp \bra a_x^\dag(t) a_0(0)\ket_{\lambda;\xi}\,
	\big\bra  \re^{\lambda\Omega_0(x,t)} \big\ket
\eeq
where $\bra\cdots\ket_{\lambda;\xi}$ is the state characterised by $\beta^j(\lambda;\xi)$ as determined by \eqref{flow}. Recall that $\xi = x/t$, and that for the free-fermionic model we are considering, one uses the function $w(k)$ instead of the Lagrange parameters $\beta^j$, and on the flow this has the explicit solution $w(\lambda;\xi;k)$ given by \eqref{floww} with $h_{i_*}(k) = h_0(k) = 1$.

\begin{figure}[ht]
	\begin{center}
		\includegraphics[width=0.3\textwidth]{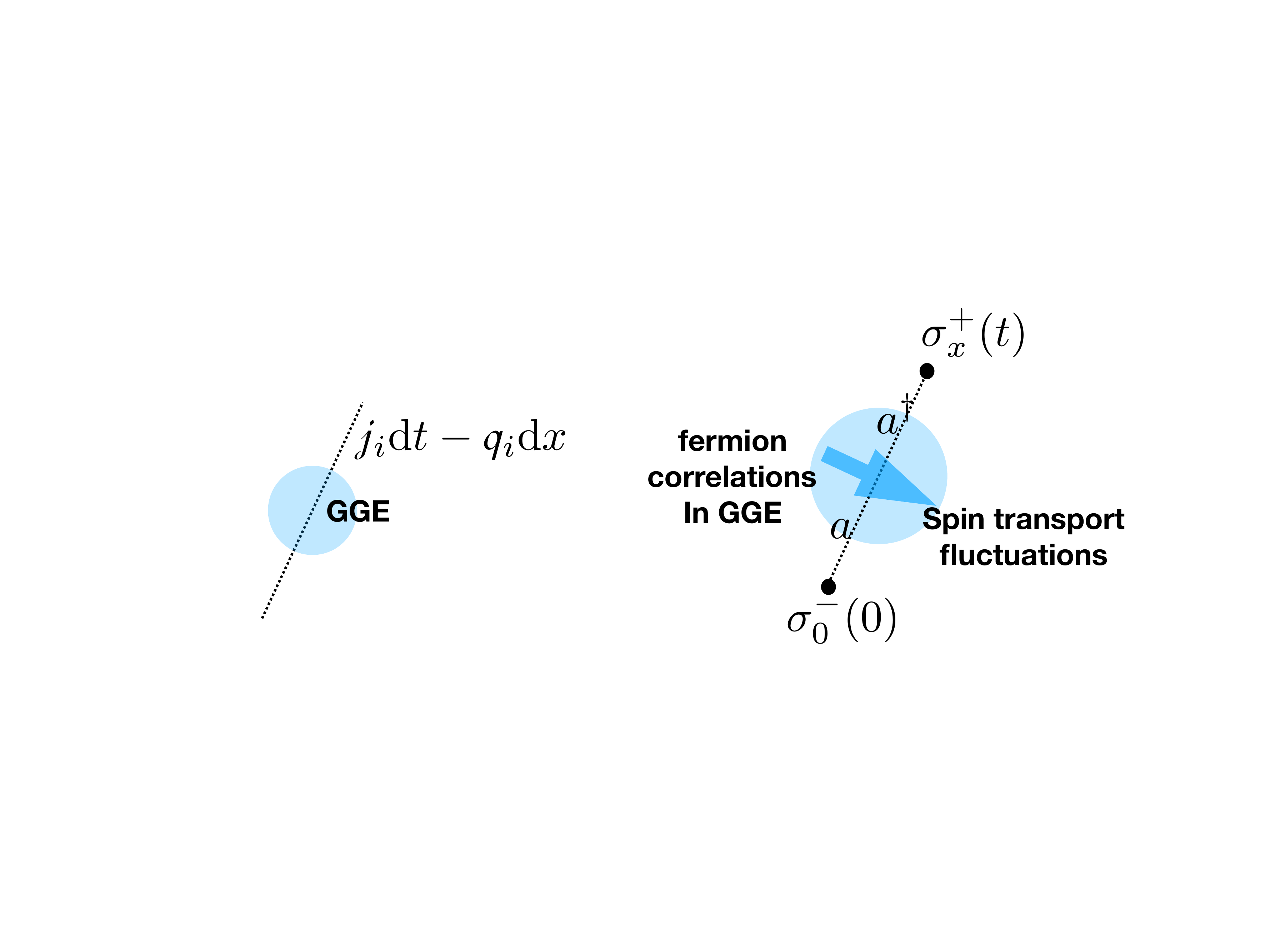}
	\end{center}
	\caption{The leading exponential decay of the two-point function \eqref{corrOmega} is controlled by two effects, as expressed in \eqref{corraa}: the (possible) exponential decay of the fermion correlation within the state modified by the presence of the JW string (represented by the blue shaded region), and the contribution due to the space-time free energy representing the spin fluctuations and given by the BFT (represented by the blue arrow).}
	\label{fig:sigmafleche}
\end{figure}

The asymptotic \eqref{corraa} is argued for by using Wick theorem in Appendix \ref{appassfacto} (only the case $t=0$ is analysed for simplicity) and it is supported by an \emph{exact} factorisation of the correlation function found in the literature (see \eqref{eq:its_factorisation}) and possible one-to-one correspondence of the terms involved will be subject of future investigations (see Appendix \ref{comparison_its}). The physical idea behind it is that the leading exponential behaviour of the correlator on the left-hand side of \eqref{corraa} is controlled by two effects. The first is the (eventual) exponential decay due to the interaction between the fermions $a^\dag_x(t)$ and $a_0(0)$, which occurs in the region ``deep" between $(0,0)$ and $(x,t)$ and is thus in a state modified by the presence of the space-time JW string; the second is the exponential decay due to the change of ``free energy" induced by the space-time JW string representing the spin fluctuations, as given by the BFT, a classical hydrodynamical effect. See Fig.~\ref{fig:sigmafleche}. In section \ref{ssectexpo} we use \eqref{corraa} in order to obtain the explicit exponential decay of the transverse correlation function.

\subsection{Exponential behaviour of transverse correlation function}\label{ssectexpo}

We now analyse the leading exponential behaviour as obtained from \eqref{corraa}. Our main interest is the exponential decay of correlation functions. In principle, our methods, including the BFT, provide also certain oscillations on top of the decay. Although this is an interesting subject, we still don't have a full understanding, and thus we focus on the exponentially decaying envelope. We hope to return to oscillations in future publications.

The first contribution to \eqref{corraa} comes from the two-point function of fermions. This is obtained by Wick's theorem, and using \eqref{floww} we get
\begin{equation}
	\bra a^\dag_x(t) a_0(0 )\ket_{\lambda;\xi} = \int_{-\pi}^\pi\frac{\ud k }{2\pi}\frac{\e^{\ri E(k)t-\ri kx}}{1+\e^{w(k) + \lambda\, {\rm sgn}(x- v(k)t)}}.
	\label{eq:shifted_w}
\end{equation}
For its application to Eq.~\eqref{corraa}, the only relevant part is the asymptotic exponential behaviour of \eqref{eq:shifted_w}. In the time-like region $|\xi|< 4$ the phase has a stationary point on the integration contour, and the behaviour is algebraic for all values of $\lambda$. The same conclusion holds for $|\xi|=4$, thus
\begin{equation}
	\bra a^\dag_x(t) a_0(0 )\ket_{\lambda;\xi} \asymp 1\qquad(|\xi|\leq4).
	\label{eq:shifted_w_time}
\end{equation}

In the space-like region no stationary point exist on the integration contour, and the exponential behaviour is obtained by an application of contour deformations and the steepest descent method. The formula simplifies as in this region, ${\rm sgn}(x-  v(k)t) = {\rm sgn}(x)$. For evaluating \eqref{corrOmega}, we need $\lambda = \ri\pi$, and thus
\begin{equation}
	\bra a^\dag_x(t) a_0(0 )\ket_{\ri\pi;\xi} = \int_{-\pi}^\pi\frac{\ud k }{2\pi}\frac{\e^{\ri E(k)t-\ri kx}}{1-\e^{w(k)}}\qquad (|\xi|>4).
	\label{eq:shifted_w_space}
\end{equation}

We obtain the following estimate (see Appendix \ref{appsaddle}):
\beqa
	\bra a^\dag_x(t) a_0(0 )\ket_{\ri\pi;\xi} &\asymp& \exp\big(-\arccosh(\xi/4)\, |x| + \sqrt{x^2-16t^2} - \ri \pi |x|/2 + 2\ri h \,t \big)
	+ \Lambda \n  &=&
	\exp\big(-M_\xi\, |x|  - \ri \pi |x|/2 + 2\ri h \,t \big) + \Lambda 
	\qquad(|\xi|>4)
	\label{exponentialbehaviour}
\eeqa
where
\beq\label{Mxi}
	 M_\xi = \arccosh(\xi/4)-\sqrt{1-\frc{16}{|\xi|^2}}
\eeq
and $\Lambda$ is the contribution coming from residues of the integrand due to the poles in the denominator (again, see Appendix \ref{appsaddle}).
In \eqref{exponentialbehaviour}, the most slowly decaying exponential amongst the terms on the right-hand side is to be taken.

It is worth analysing this for the thermal state, $w(k) = \beta E(k)$. In the space-like region $|\xi|>4$, the quantity $\Lambda$ is obtained from solving $w(z)=0$. The solution depends on the value of $|h|$.
\bi
\item If $|h|\leq 2$, the gapless regime, then a zero exists at $z=0$ and we find $\Lambda = 1$ and therefore
\beq
	\bra a^\dag_x(t) a_0(0 )\ket_{\ri\pi;\xi} \asymp 1.
\eeq
Thus in the gapless regime, the two-point function of fermions does not give additional exponential decay in \eqref{corrOmega}, and the exponential behaviour of the transverse correlator is fully determined by spin fluctuations.
\item If $|h|>2$, there is a nontrivial solution with $\hat z=\hat k + \ri \hat q$ with $\hat k=0$ ($\hat k=\pi$) for $h>0$ ($h<0$), and with $\hat q=\arccosh(h/2)$. This contributes to $\Lambda$ if and only if $2|h|<|\xi|$, in which case it is the leading contribution to $\Lambda$ and otherwise $\Lambda=1$, thus
\beq
	\Lambda = \exp \big(-\arccosh(h/2) \,|x|\big)\times
	\lt\{\ba{ll} (-1)^x & (h<0) \\ 1 & (h>0).\ea\rt.
\eeq
Therefore we find
\beq
	\bra a^\dag_x(t) a_0(0 )\ket_{\ri\pi;\xi}\asymp
	\re^{\ri \Phi}\exp\big(-{\rm min}\big(\arccosh(h/2),  M_\xi\,\big)\,|x|\,\big)\label{finalgapped}
\eeq
where
\beq
	\Phi =\lt\{\ba{ll} \pi x & \mbox{(first argument, $h<0$)}\\
	0 & \mbox{(first argument, $h>0$)}\\
	-\pi |x|/2 + 2 h t & \mbox{(second argument)}
	\ea\rt.
\eeq
where ``argument" refers to that taken by the min function in \eqref{finalgapped}. Thus when there is a gap between the energy of the ground state and that of the lowest excited state, an extra contribution to the exponential decay of the transverse correlator arises from the fermion correlations. It may be surprising that the gap affects the correlation function in a finite-density GGE, far from the ground state; the important point here is that the state is modified by the presence of the JW string. This transforms the distribution into that of bosons instead of fermions, which is strongly affected by the presence or not of a gap.
\ei

The second contribution in \eqref{corraa} comes from the spin fluctuations and may be evaluated using the BFT results \eqref{gasymp} with \eqref{Flambda} and \eqref{floww}. This gives
\beq
	\big\bra  \re^{\lambda\Omega_0(x,t)} \big\ket
	\asymp
	\exp\lt[\int_{-\pi}^\pi \frc{\dd k}{2\pi}\,
	|x-v(k)t |
	\log \Big(\frc{1+e^{-w(k)- \lambda\, {\rm sgn}(x- v(k)t)}}{1+e^{-w(k)}}\Big)
	\rt].
\eeq
We must evaluate this at $\lambda = \pm\ri \pi$. The sign here does not affect any of the calculations, however the direction in which the analytic continuation of the explicit asymptotic formula is taken affects the result. We believe that this is because the analytic continuation to $\pm\ri \pi$ does not commute with the evaluation of the asymptotics. However, the exponent in the asymptotic is only affected by a purely imaginary number; this contributes to oscillations, which we do not address here. The exponential decay of the absolute value of the average is not affected by the direction of the analytic continuation, and we obtain
\begin{equation}
	\left|\big\bra  \re^{\ri\pi\Omega_0(x,t)} \big\ket\right|
	\asymp
	\re^{f_{x,t}[w]}\quad.
\end{equation}
where we use the notation $f_{x,t}[w] = f_{\ri\pi,x,t}[w]$ from \eqref{Flambda}, giving
\begin{equation}
	f_{x,t}[w] = \int_{-\pi}^\pi \frc{\dd k}{2\pi}\,
	\lt|x-v(k)t \rt|
	\log \Big|\tanh\frc{w(k)}2\Big|\quad .
	\label{eq:f_functional}
\end{equation}
In Appendix \ref{appsaddle} we write a general expression for the transverse correlation function valid on arbitrary homogenous GGEs, see \eqref{fullresulttransverse}. For the thermal state, $w(k)=\beta E(k)$, a further simplification occurs in the space-like region: using the fact that $v(k)$ is odd and that $E(k)$ is even we find the form presented in \eqref{eq:therm_F} in the introduction.

\subsection{Comparison with numerical results}

\begin{figure}[ht]
	\begin{center}
		\includegraphics[width=0.42\textwidth]{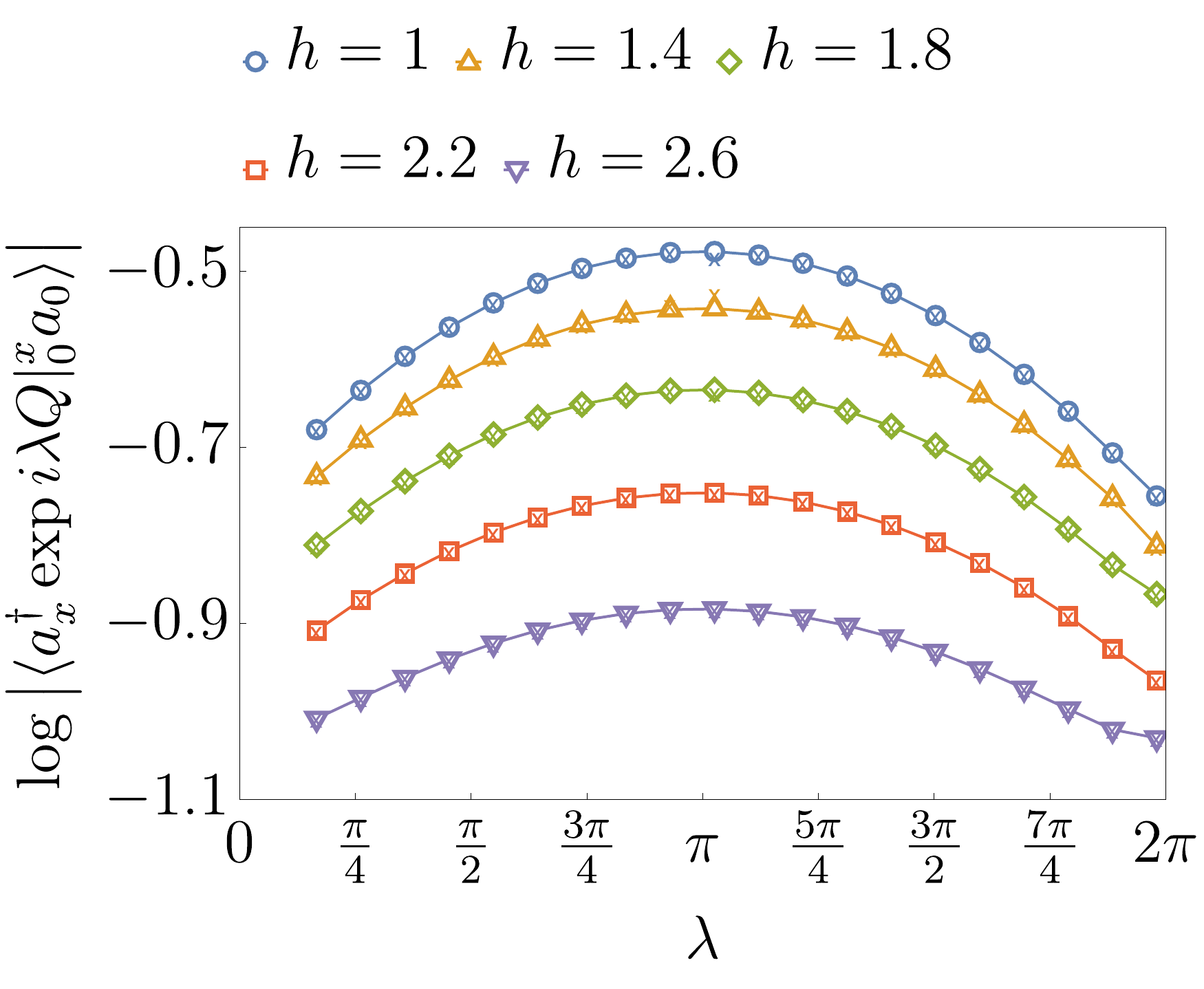}
		\includegraphics[width=0.42\textwidth]{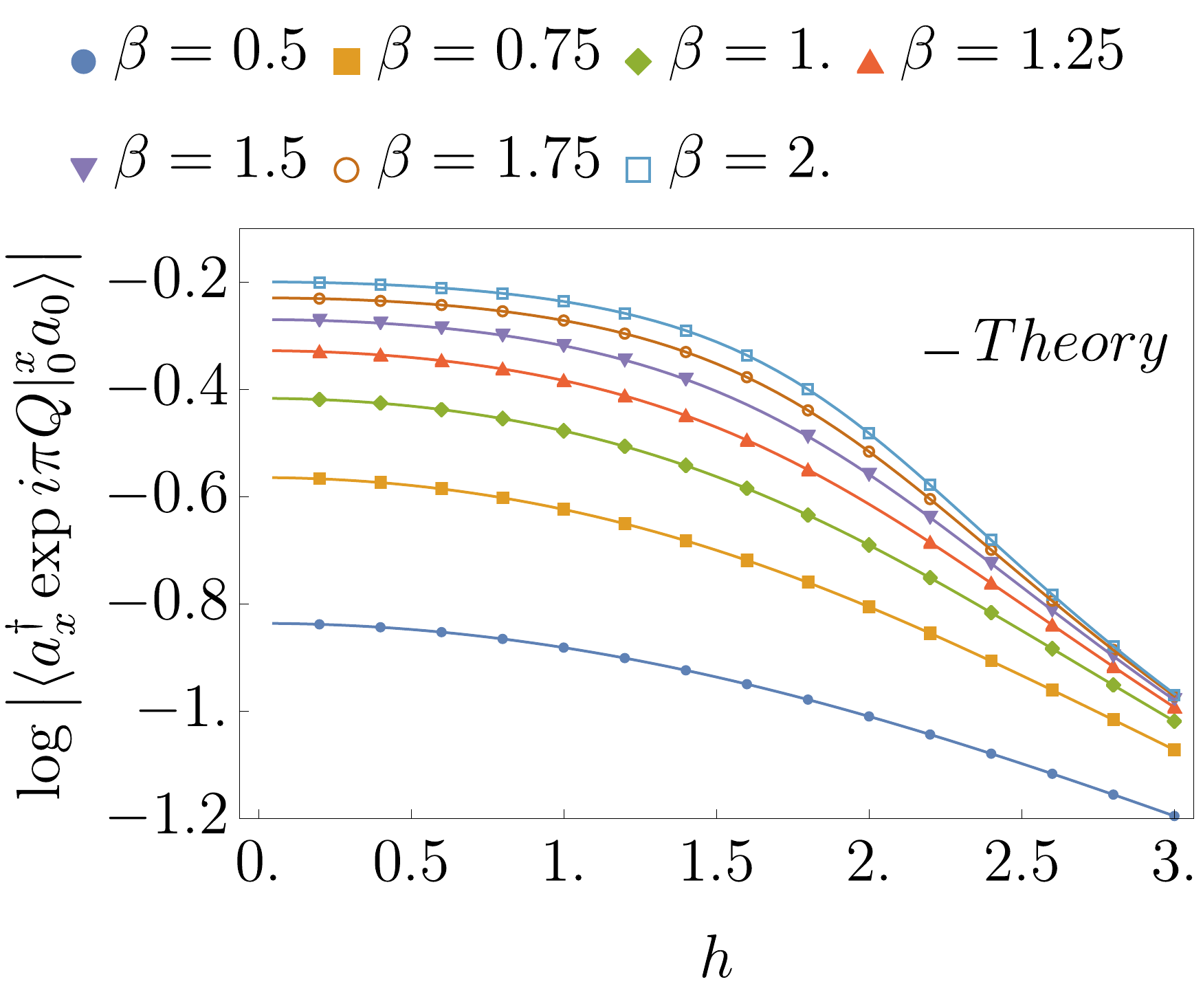}
	\end{center}
	\caption{(Left) Numerical evidence of the factorisation formula in eq. \eqref{corraa} at $t=0$ for a spin chain $N=150$ long on a thermal state at inverse temperature $\beta=1$. Markers joined by the continuous line represent the data points obtained numerically computing the left hand side of eq. \eqref{corraa} while the crosses represent the same quantity computed using the factorisation Ansatz on the right hand side of the same equation. (Right) Exponent governing the decay with distance of the two-point spin-spin transverse correlation function eq. \eqref{transverse} on thermal states in the purely space-like regime $t=0$ for different values of the temperature and magnetic field. The continuous lines represent the theoretical prediction obtained from \eqref{eq:therm_F} while the markers represent numerical evaluation according to procedures described in Appendix \ref{app:numerical_methods}.}
	\label{fig:space_like}
\end{figure}

\begin{figure}[ht]
	\begin{center}
		\includegraphics[width=0.42\textwidth]{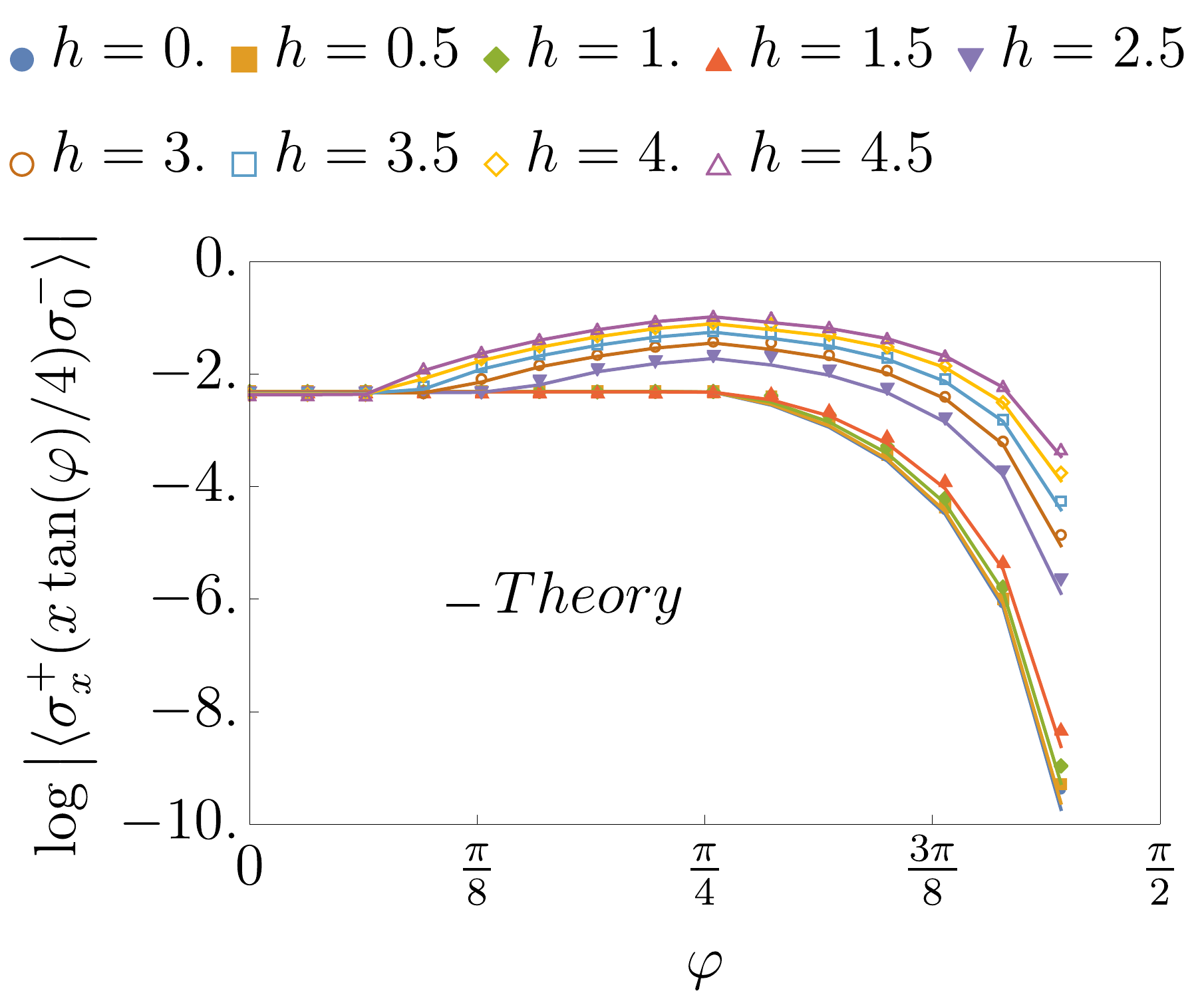}
		\includegraphics[width=0.42\textwidth]{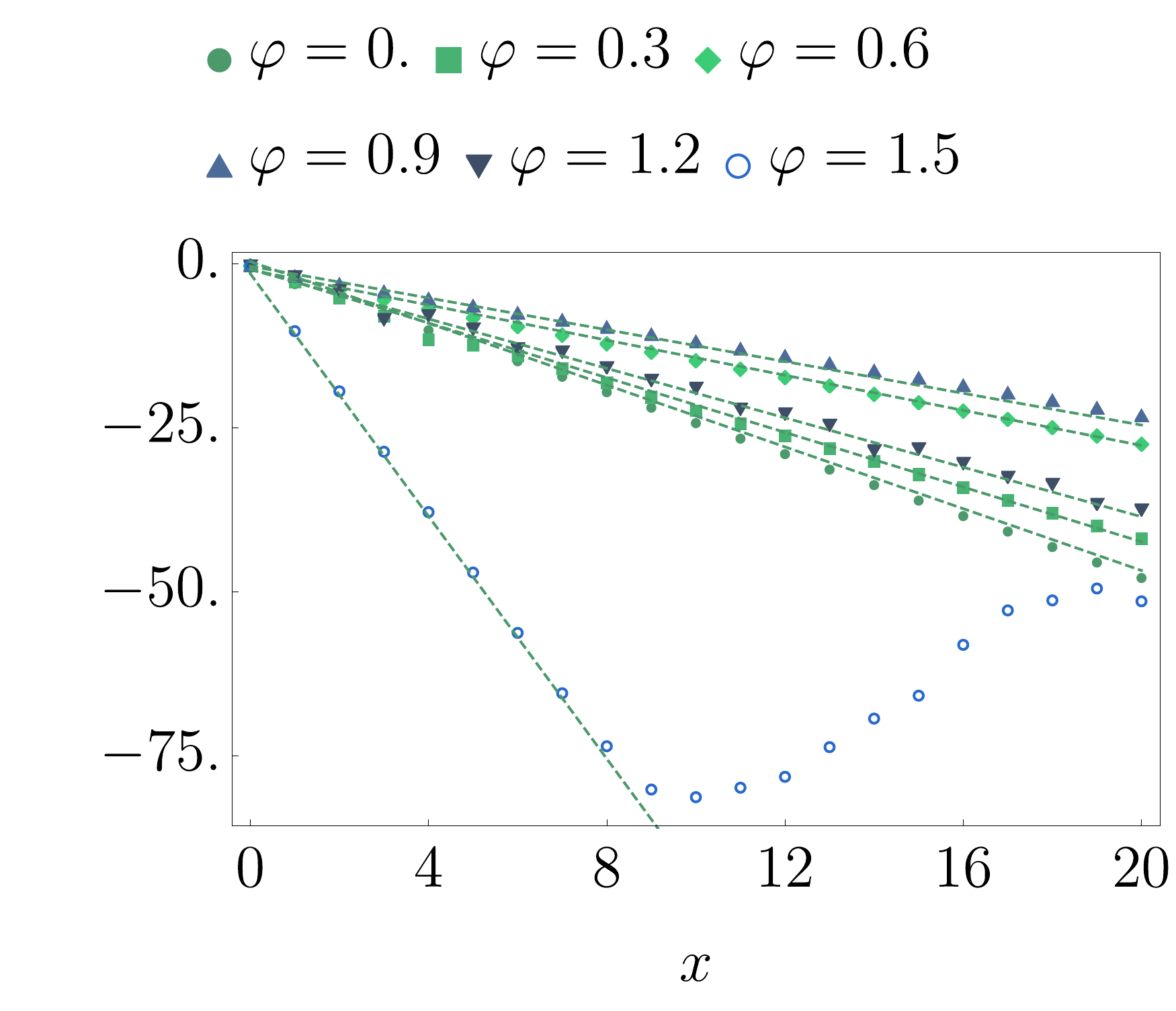}
	\end{center}
	\caption{Exponential decay of dynamical transverse two-point function on a thermal state at inverse temperature $\beta=0.1$. (Left) Comparison between analytical prediction for the exponential decay along rays eq. \eqref{eq:prediction_rays} (continuous line) as function of the ray angle $\varphi$ and numerics (markers). We plot different curves each for a certain value of the magnetic field in both the massless ($|h|>2$) and the massive regime ($|h|\leq 2$). Analytically predicted values are shown as crosses joined by segments. (Right) Decay of the correlation function for $h=4$ along the ray as function of the spatial variable $x$ together with the theoretically predicted slopes.}
	\label{fig:dynamical}
\end{figure}

In this section we show and discuss how the analytic predictions for the transverse correlation function obtained in the previous sections compare with numerical results. There are a number of studies in the literature that considered the time evolution of correlation functions in spin chain \cite{PhysRevB.84.052406, PhysRevB.53.8486, PhysRevB.52.4319}, however the numerical results we present are new.
In Fig. \ref{fig:space_like} we show that the factorisation ansatz in \eqref{corraa} is extremely accurate and works very well in wide regions of parameters space. Here, we have considered the static spin-spin transverse correlation function in eq. \eqref{transverse} on a thermal state at fixed temperature $\beta=1$ for different magnetic fields $h$. In the left panel, we checked the factorisation formula for $\lambda\mapsto-\ri\lambda$ purely imaginary around the physically relevant point $\lambda=\pi$. More specifically, on one hand, we numerically compute the left hand side of eq. \eqref{corraa} using the methods of Appendix \ref{app:numerical_methods} and on the other hand we numerically compute the two factors on the right hand side separately. The first factor is given by the correlation function of the pure twist field $\vev{e^{\ri\lambda Q|_0^x}}$ (again computed following Appendix \ref{app:numerical_methods}) while the second is the integral given in \eqref{eq:shifted_w} with $w(k)=\beta E(k)$. On the right, we choose $\lambda=\pi$ and study the behaviour of the correlation function varying the parameters $\beta$ and $h$ going from the massive to the massless phases. Note that the exponent appears to be smooth, even though the explicit form of the formula depends strongly on if $|h|>2$ or $|h|<2$ in the cases $|\xi|>4$. The numerical data points are obtained as explained in the steps below, with the choice $\varphi=0$ that is the purely spacial regime (non-dynamical). In Fig. \ref{fig:dynamical} we focus instead on the fully dynamical transverse correlation function on a thermal state again. We choose $\beta=0.1$ for numerical simplicity but we explore both the massive and massless phases of the model. We note that in the literature the massive regime has been consistently overlooked and this is the first time that the correlation function decay behaviour is explored going \emph{along} the ray: previous studies focused either on purely spatial or on purely dynamical regimes. It is also the first time the region of parameters $|\xi|>4$ (space-like) and $|h|>2$ (massive phase) is explored. Note that a ray can be parametrised by an angle as $\frac{x}{4t}=\frac{\xi}{4}=\cot\varphi$. To obtain data we proceed as follows:
\begin{itemize}
	\item We fix a ray by choosing a value for $\varphi$.
	\item We fix a value of the magnetic field $h$ (temperature is fixed throughout to $\beta=0.1$).
	\item We compute (according to Appendix \ref{app:numerical_methods}) the spin-spin transverse correlation function \eqref{transverse} along the chosen ray varying $x\in[0,20]\subset\Z$ and parametrising time as $t=\frac{x\tan\varphi}{4}$.
	\item We fit the exponent of the numerically obtained data (see right panel of Fig. \ref{fig:dynamical}).
	\item We repeat the procedure for various $\varphi$'s and $h$'s, and $\lambda$'s in the static case (Fig,~\ref{fig:space_like}), and plot the resulting exponents.
\end{itemize}
In the left panel of Fig. \ref{fig:dynamical} the theoretical prediction of eq. \eqref{eq:therm_F} is indicated by the continuous line and fits well with the numerical results given by the markers. In particular, along a ray we have
\beqa\label{eq:prediction_rays}
\lefteqn{F\lt(x, \frac{x\tan\varphi}{4}\rt)/|x|}\\&&=\lt\{
\begin{aligned}
	&\int_{-\pi}^\pi\frac{\dd k}{2\pi}\lt|1-\frac{v(k)\tan\varphi}{4}\rt|\log\lt|\tan\lt(\frac{\beta E(k)}{2}\rt)\rt| &&(\frac{\pi}{4}<\varphi<\frac{\pi}{2})\\ 
	& \int_{-\pi}^\pi\frac{\dd k}{2\pi}\log\lt|\tan\lt(\frac{\beta E(k)}{2}\rt)\rt| &&(0\leq \varphi \leq \frac\pi4,\ |h|\leq 2)\\
	& -{\rm min}\big(\arccosh(h/2),  M_\xi\big)+\int_{-\pi}^\pi\frac{\dd k}{2\pi}\log\lt|\tan\lt(\frac{\beta E(k)}{2}\rt)\rt|
	&&(0\leq \varphi \leq \frac\pi4,\ |h|>2)
\end{aligned}
\rt.\no\quad .
\eeqa
On the right panel we have the same correlation function from a different perspective: we plot it as a function of the spatial variable $x$ for different rays at fixed magnetic field $h=4$. The dashed lines are obtained predicting the slopes using formula \eqref{eq:prediction_rays} and fitting the intercept from data.

A remark is to be done: we can notice that on the ray given by $\varphi=1.5\approx\pi/2$ the exponential decay behaviour is broken around $x=8$. This happens because quasiparticles responsible for the decay of the correlation function have a maximal velocity and may eventually hit the boundary of the chain when travelling for long time. If the maximal velocity is $v_{\rm max}$ and the chain length is $N$ at time $t_{\rm max}=\frac{N}{v_{\rm max}}$ the fastest excitation will have hit the boundary and will bounce back causing and will scatter with other quasiparticles modifying the behaviour of the correlation function. In our simulation the chain was made by $N=120$ spins and the maximal velocity in the XX model with our parametrisation is $v_{\rm max}=4$ so that $t_{\rm max}=30$ and along the ray with $\varphi=1.5$ we obtain $x_{\rm max} = 4 t_{\rm max}\cot(1.5)\approx 8$ confirming our numerical findings.

\section{Conclusion} \label{conclu}

In this paper we have analysed the correlation functions of the XX model using hydrodynamic techniques. This served to illustrate two different hydrodynamic theories for asymptotic of correlation functions in space time: the hydrodynamic projection theory for algebraic decay of generic fields (hydrodynamic linear response theory), and the ballistic fluctuation theory (BFT) for exponential decay of twist fields.

The former theory is well established, and we verified its validity for longitudinal correlatiors of the XX model. Importantly, this emphasised the importance of fluid-cell averaging for the hydrodynamic results to hold, and made explicit the definition of an appropriate fluid-cell mean in the case of this correlator. 

The latter theory was proposed in \cite{doyon2019fluctuations}, but never since explored in explicit models. We analyse the transverse correlators of the XX model, which can be seen as twist fields, and provide the first confirmation of the theory. For this purpose, two new results are established.

First, the application of the BFT to transverse spin correlator $\bra \sigma_x^+(t)\sigma_0^-(0)$ requires the consideration of ``fermionic descendants" of twist fields: by the Jordan-Wigner transformation, local fermion creation and annihilation operators are present at the positions of the twist fields, at the end-points of the Jordan-Wigner string. We explain how a factorisation formula allows us to separate the contribution of the fermion correlator, from the contribution of the Jordan-Wigner string average. Both have different interpretations: the fermion correlator contributes a possible exponential decay equal to that obtained in the original GGE modified by the presence of the JW string. The Jordan-Wigner string average, obtained by the BFT, represents the fluctuations of the spin 2-current through the ray connecting the points $(0,0)$ to $(x,t)$. The latter is a full hydrodynamic theory for the physical idea of counting domain walls, first introduced by Sachdev et al. \cite{Sachdev_1996, Sachdev_1997, Buragohain_1999}.

Second, we have shown how the Jordan-Wigner string contribution to the dynamical correlation function can be written as a ``space-time Jordan-Wigner string". Two forms of the space-time Jordan-Wigner string are provided: in the form \eqref{corrOmegaproduct}, the result is expected to be an exact operator equation, while in the form \eqref{corrOmega}, it is an asymptotic result for the average of the string. In the latter, the space-time Jordan Wigner string is an exponential of the integrated perpendicular 2-current (in the discrete-continuous space-time) from the start to the end point. This is a generalisation of the usual Jordan-Wigner string to the space-time domain. This is, in a sense, the real-time equivalent of twist fields considered in statistical models \cite{ZuberItzykson,SchroerTruong}. The BFT gives exact asymptotics for averages of space-time Jordan-Wigner strings.

Our techniques are expected to easily generalise to the XY chain, using again its free fermion structure. In the literature, a conjecture for the XY model already exists  \cite{PhysRevB.84.052406}. In that work the authors only consider the parameter range $|h|<2$ (in our normalisation), for which, when specialised to the XX model, they recover the formula of Its et.\ al.\ \cite{its1993temperature}. As we saw, in this regime the only contribution to the asymptotic exponential decay in the XX model comes from the space-time Jordan-Wigner string, because for $|h|\leq 2$ the model is gapless. But, as we also saw, when a gap is present, $|h|>2$, the fermion correlators provide an additional contribution in the space-like region, see the result \eqref{eq:prediction_rays}. We expect the same phenomenon to occur more generally in the XY model. That is, we expect the conjecture of \cite{PhysRevB.84.052406} to be modified by additional contributions from fermion correlators in the space-like region, in any parameter regime of the XY model where the model is gapped. The XY model is generically gapped even for $|h|<2$, and thus the conjecture of \cite{PhysRevB.84.052406} is likely to be incomplete. We will come back to this in a further work.

We stress that we expect many of our findings to hold in other more generally interacting models with local interactions on the basis of general principles of many-body physics, including exponential clustering in space of thermal correlation functions \cite{Araki}  and Lieb-Robinson bounds \cite{LiebRobinson}.

There is a number of avenues which this work opens. Other twist fields in free fermion models are immediately accessible, such as spin operators in the XY model as mentioned above. As the theory is based on hydrodynamics, it is also applicable to interacting models. There, the difficulty may be to correctly identify the twist fields, and account for possible local-field descendants of twist fields and the resulting factorisation -- our technical argument for factorisation was based on the free-fermion structure, but the physical intuition seems to be more general. Interacting models of interest are the Lieb-Liniger and Sine-Gordon models, where the fundamental field, and the vertex operators, respectively, may be accessible by the BFT. Other twist fields of importance are the branch-point twist fields studied in the context of entanglement entropy; we believe our theory can be adapted to obtain new results for asymptotics of entanglement entropy out of equilibrium. It would be interesting to extend the BFT in order to access diffusive corrections, which might help give information about algebraic pre-factors to the exponential decay. It would also be interesting to understand oscillatory behaviours. Although we concentrated on two-point functions, we believe simple arguments will give access to large-time behaviours of one-point function of twist fields in quantum quenches. On the mathematical side, in free-fermion models, the objects we are considering, including averages of Jordan-Wigner strings, can be seen as Fredholm determinants, and our techniques offer a way, based on hydrodynamic ideas, of obtaining their asymptotics especially in relation with non-linear steepest descent and Riemann-Hilbert factorisation problems (see Appendix \ref{comparison_its}). Finally, the BFT is a particular construction for evaluating large-deviations of extensive quantities out of equilibrium. Recent work shows that the macroscopic fluctuation theory, traditionally used in diffusive systems, can be adapted to the ballistic regime, and this could be useful for studying twist fields following the ideas developed here.

{\bf Acknowledgment}: We would like to thank Denis Bernard for important discussions at the beginning of this projects. We also thank Frank G\"ohmann, Pierre Le Doussal, Nikolay Kitanine, M\'arton Kormos and Joachim Stolze for useful comments. BD is  supported by EPSRC under the grant ``Emergence of hydrodynamics in many-body systems: new rigorous avenues from functional analysis", ref.~EP/W000458/1. BD's research was also supported in part by the International Centre for Theoretical Sciences (ICTS) for the online program - Hydrodynamics and fluctuations - microscopic approaches in condensed matter systems (code: ICTS/hydro2021/9).

\begin{appendices}

\section{Fredholm determinants, integrable PDEs and Euler Hydrodynamics}\label{comparison_its}

In this Appendix we recall the basic ingredients which traditionally led to the calculation of explicit asymptotic of certain correlation functions of quantum integrable models, in particular the XX model considered here. The basic idea that allowed the extraction of the asymptotic has been the observation that quantum correlation functions can be expressed in terms of quantities that satisfy classical non-linear integrable PDEs. In turn, these PDEs can be equivalently formulated as Riemann-Hilbert factorisation problem for which expansions of the solution for large values of the parameter is possible due to special analytic properties of the conjugation matrix. For more details see the book \cite{korepin_bogoliubov_izergin_1993} and references therein. The theory of integrable PDEs is a well estabilished subject of mathematical physics with connections to the theory of integrable (Fredholm) operators, Painlev\'e equations, hydrodynamics, random matrix theory and much more.

In the XX model case considered here the transverse correlation function is expressed exactly as a product \cite{its1993temperature}
\begin{equation}\label{eq:its_factorisation}
	\vev{\sigma^+_x(t)\sigma^-_0(0)} = \e^{-2iht} b_{++}(x,t) \e^{\sigma(x,t)}\quad .
\end{equation}
Let us explain the various terms here. Let $\lambda$ and $\mu$ be complex spectral parameters on the unit circle $|\lambda|=|\mu| = 1$. Then, $\sigma(x,t)$ is the logarithm of a Fredholm determinant associated with the integrable operator having the following kernel
\begin{equation}
	V(\lambda,\mu) = \frac{e_+(\lambda) e_-(\mu) - e_-(\lambda) e_+(\mu)}{\lambda - \mu}
\end{equation}
and it is explicitely written as follows
\begin{equation}
	\sigma(x,t) = \log\det\left(\hat{1} + \hat{V}\right)\quad.
\end{equation}
The functions $e_\pm$ depend on space-time $x,t$, on (inverse) temperature $\beta$, magnetic field $h$ and the spectral parameter $\lambda$ and have the following simple form
\begin{subequations}
	\begin{equation}
		e_-(\lambda) = \sqrt{n(\lambda)}\, \lambda^{x/2}\e^{-ith(\lambda+1/\lambda)}
	\end{equation}
\begin{equation}
	e_+(\lambda) = e_-(\lambda) E(x,t,\lambda)
\end{equation}
\end{subequations}
with ($\rm{P.V.}$ is the Cauchy principal value)
\begin{equation}
	E(x,t,\lambda) = {\rm{P.V.}}\int \frac{\dd\mu}{\pi}\frac{\e^{2ith(\mu+1/\mu)}\mu^x}{\mu-\lambda}\quad .
\end{equation}
Above, $n(\lambda)$ is the occupation function appearing in all Thermodynamic Bethe Ansatz solvable models that we introduced in \eqref{eq:occupation_GGE}. The difference here is only in the parametrisation of the spectral parameter, here being $\lambda(k) = \e^{i k}$, and in the choice of a thermal state corresponding to $w(k) = \beta E(k)$, proportional to the dispersion relation as in \eqref{thermal}. The ``potential" $b_{++}(x,t)$ can be shown to satisfy the discretised non-linear Schr\"odinger equation \cite{https://doi.org/10.1002/sapm1976553213}. In turn, space-time derivatives of $\sigma(x,t)$ can be written in terms of the $b_{++}(x,t)$. After, rewriting the integrable PDEs system mentioned above as a Riemann-Hilbert matrix factorisation problem one can extract the asymptotic in relevant regimes. We believe our BFT formalism can have connections with non-linear steepest descent method \cite{10.2307/2946540}.
\section{Asymptotic factorisation} \label{appassfacto}

In this section we consider the factorisation property \eqref{corraa} of the transverse correlation function. Below, we give an argument supporting its validity that will complement the numerical evidence provided in Fig.~\ref{fig:space_like}. First we write
\begin{equation}\label{firsteq}
	\vev{\re^{\lambda Q|_y^x}a^\dag_x a_y} = \frac{\vev{\re^{\lambda Q|_y^x}a^\dag_xa_y}}{\vev{\re^{\lambda Q|_y^x}}}\vev{\re^{\lambda Q|_y^x}}=:\vev{a^\dag_x a_y}_\lambda\vev{\re^{\lambda Q|_y^x}}
\end{equation}
Recall that $Q|_y^x = \sum_{z=y}^{x-1} q(z)$ and that $q(z) = a^\dag_z a_z$; for simplicity, we assume $x>y$. We have defined $\vev{\dots}_\lambda = \vev{\re^{\lambda Q|_y^x}\dots}/\vev{\re^{\lambda Q|_y^x}}$. The transverse correlation function is obtained for $\lambda=\ri\pi$.

Let us now study the following static $\lambda$-dependent expectation value
\beq
	g_\lambda(z,z')= \vev{a^\dag_z a_{z'}}_\lambda
\eeq
for $z,z'\in[y,x]$. First, by differentiating we get
\begin{equation}
	\p_\lambda g_\lambda(x,y) = \vev{ Q|^x_y a^\dag_x a_y}_\lambda - \vev{a^\dag_x a_y}_\lambda\vev{ Q|^x_y}_\lambda .
\end{equation}
Using Wick theorem\footnote{Wick's theorem holds because $ Q|_y^x$ is a quadratic form (is quadratic in fermions and preserve fermion number). By the Backer-Campbell-Hausdorff formula $\re^A\re^B=\re^{A+B+C}$ where $C$ is formed of nested commutators of $A$ and $B$. The subspace of the fermionic algebra made by quadratic forms is a subalgebra with respect to the product given by the commutator. It follows that also the series of nested commutators $C$ is an element of the algebra, and so a quadratic form.} we compute
\begin{equation}
	\vev{ Q|^x_y a^\dag_x a_y}_\lambda = \vev{ Q|^x_y}_\lambda\vev{a^\dag_x a_y}_\lambda - \sum_{z=y}^{x-1} \vev{a^\dag_z a_y}_\lambda\vev{a^\dag_x a_z}_\lambda
\end{equation}
which gives
\begin{equation}
	\p_\lambda g_\lambda(x,y) = - \sum_{z=y}^{x-1} \vev{a^\dag_z a_y}_\lambda\vev{a^\dag_x a_z}_\lambda = -\sum_{z=y}^{x-1}g_\lambda(x,z) g_\lambda(z,y).\label{eq:g_alpha}
\end{equation}
Likewise,
\begin{equation}
	\p_\lambda g_\lambda(x,z) = -\sum_{z'=y}^{x-1}g_\lambda(x,z') g_\lambda(z',z) .\label{eq:g_alphazz}
\end{equation}

For simplicity, we assume parity symmetry of the state; the argument can be generalised to states which are not parity symmetric. Let us assume that $g_\lambda(x,y)$, $g_\lambda(x,z)$, $g_\lambda(z,y)$ and $g_\lambda(z',z)$ decay exponentially at large separations:
\beq\begin{aligned}
	g_\lambda(x,y) &\asymp e^{-u_\lambda (x-y)} && (x\gg y)\\
	g_\lambda(x,z) &\asymp e^{-v_\lambda (x-z)} && 
	(x\gg z \gg y)\\
	g_\lambda(z,y) &\asymp e^{-v_\lambda (z-y)} && 
	(x\gg z \gg y)\\
	g_\lambda(z',z) &\asymp e^{-w_\lambda |z'-z|} && 
	(x\gg z,z' \gg y,\ |z-z'|\to\infty)\\
	\end{aligned}
\eeq
for some $u_\lambda,\,v_\lambda,\,w_\lambda\geq 0$. Note that by our assumption of parity symmetry $g_\lambda(x,z)$ and $g_\lambda(z,y)$ have the same asymptotic behaviour. Near the boundaries of the asymptotic regions displayed (that is, for $z$ or $z'$ near to $x$ or $y$), the exponential behaviours are modified, but we assume that these modifications are essentially supported on finite numbers of sites. This is what is expected to happen in clustering states of local quantum chains. Then, we can use these asymptotic forms in \eqref{eq:g_alpha} and \eqref{eq:g_alphazz}, as they are satisfied for an extensive number of summands. We get
\beq
	\p_\lambda u_\lambda (x-y)\, e^{-u_\lambda (x-y)}
	\asymp (x-y) e^{-v_\lambda(x-y)}
\eeq
and
\beqa\label{abfgt}
	\p_\lambda v_\lambda(x-z) e^{-v_\lambda (x-z)}&\asymp&
	\sum_{z'=y}^{x-1} e^{-v_\lambda(x-z') - w_\lambda |z'-z|}\\
	&=& \frc{e^{-w_\lambda(x-z)} -
	e^{-v_\lambda(x-z)}}{
	e^{v_\lambda-w_\lambda} -
	1}
	+
	\frc{e^{-v_\lambda(x-z)} -
	e^{-v_\lambda(x-y)-w_\lambda(z-y)}}{
	e^{v_\lambda+w_\lambda} -
	1}\no
\eeqa
As the exponential forms must match, we must have
\beq
	u_\lambda = v_\lambda = w_\lambda.
\eeq
In particular, in \eqref{abfgt} this gives
\beq
	\p_\lambda w_\lambda(x-z) e^{-w_\lambda (x-z)}\asymp
	(x-z) e^{-w_\lambda(x-z)}
	+
	\frc{e^{-w_\lambda(x-z)} -
	e^{-w_\lambda(x-y + z-y)}}{
	e^{2w_\lambda} -
	1}
\eeq
which is consistent as $e^{-w_\lambda(x-y + z-y)} \ll e^{-w_\lambda(x-z)}$.
Finally, we can evaluate $w_\lambda$ directly: $g(z',z)$ in the region $x\gg z,z'\gg y$ is given by $\bra a^\dag_z a_z'\ket_{-\lambda;\infty}$ as defined after \eqref{corraa} (that is, at $\xi=\infty$). This is because in this region, the boundaries $y,x$ of $Q|_y^x$ are far from the positions $z,z'$ of the fermion operators, whence the insertion of $Q|_y^x$ merely modifies the GGE by the factor $e^{\lambda Q}$ (at $\xi=\infty$, this modification is indeed the result of the BFT flow equation, as reviewed in Section \ref{subsec:ballistic_fluctuation_theory}). Therefore, the quantity $w_\lambda\geq0$ is  obtained from the GGE, characterised by the function $w(k)$, as
\begin{equation}
	e^{-w_\lambda |z|} \asymp \int_{-\pi}^\pi\frac{\ud k }{2\pi}\frac{\e^{-\ri kz}}{1+\e^{w(k) - \lambda}}.\label{eq:asymptotic_approx2}
\end{equation}
(see \eqref{eq:shifted_w}). The analysis of the resulting decay can be calculated by contour deformation, see Appendix \ref{appsaddle}. We conclude that
\beq
	g_\lambda(x,y)\asymp e^{-u_\lambda(x-y)}
	= e^{-w_\lambda(x-y)} \asymp \bra a^\dag_x a_y\ket_{-\lambda;\infty}.
\eeq
With \eqref{firsteq}, this shows \eqref{corraa} in the case $\xi=\infty$.

\section{Saddle point analysis and full result for transverse correlator}\label{appsaddle}

We apply the saddle point method \cite{Bender1999} to Eq. \eqref{eq:shifted_w_space} in order to explain Eq.~\eqref{exponentialbehaviour}. For this purpose, it is sufficient to assume $|\xi|>4$.

Due to analyticity of the integrand, using Cauchy theorem we can deform the straight contour of integration $\gamma_s(r)=-\pi(1-r) + \pi r$ with $r\in[0,1]$ into any contour in the complex plane having the same end points taking care of adding contributions from residues when the new deformed contour hits a pole of the integrand. For $z=k+iq$ we have
\begin{equation}
	\bra a^\dag_x(t) a_0(0 )\ket_{\ri\pi;\xi} =\int_{\gamma_\xi}\frac{\ud z}{2\pi} \frac{\re^{t\phi_\xi(z)}}{1-\re^{w(z)}}+\Lambda
\end{equation}
where $\phi_\xi(z)=\ri(E(z)-\xi z) = q\xi - 4\sin(k)\sinh(q) + i(2h - k\xi - 4 \cos(k) \cosh(q))$, $\gamma_\xi$ is any contour starting at $z_1=-\pi$ and ending at $z_2=\pi$ and $\Lambda$ is the contribution coming from residues (its sign is not important for the evaluation of the exponential decay).  
The trick behind steepest descent is to choose $\gamma_\xi$ in such a way to have $\imag\phi_\xi=c_{\xi}\in\mathbb{R}$ constant. A simple calculation shows that these contours are also such that $\nabla\real\phi_\xi$ is higher in magnitude. This way one can bring $\re^{\ri t\imag\phi_\xi}$ out the integral and apply Laplace method because the argument of the exponential will be purely real.

\begin{figure}[h]
	\begin{center}
		\includegraphics[width=0.42\textwidth]{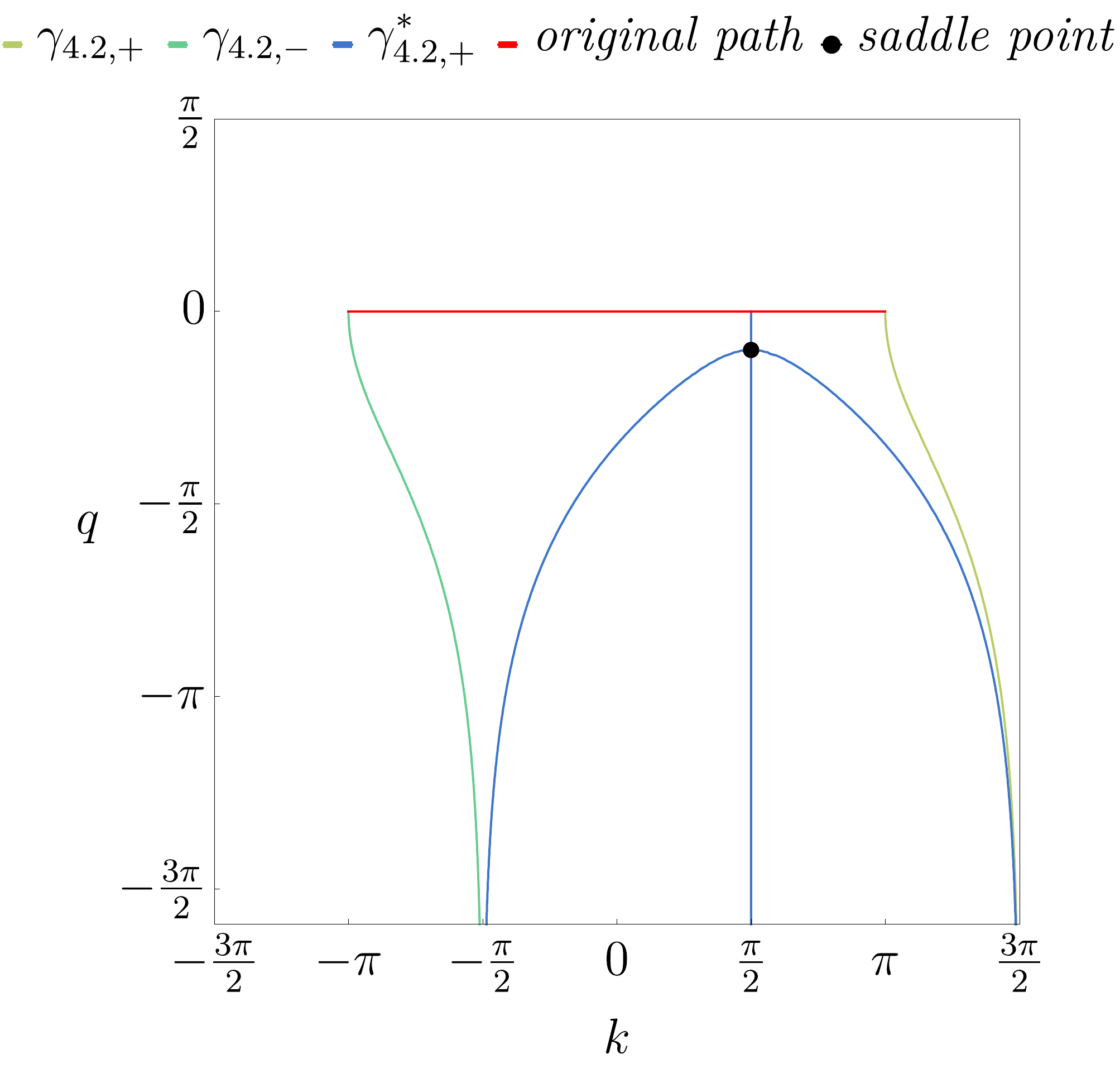}
		\includegraphics[width=0.54\textwidth]{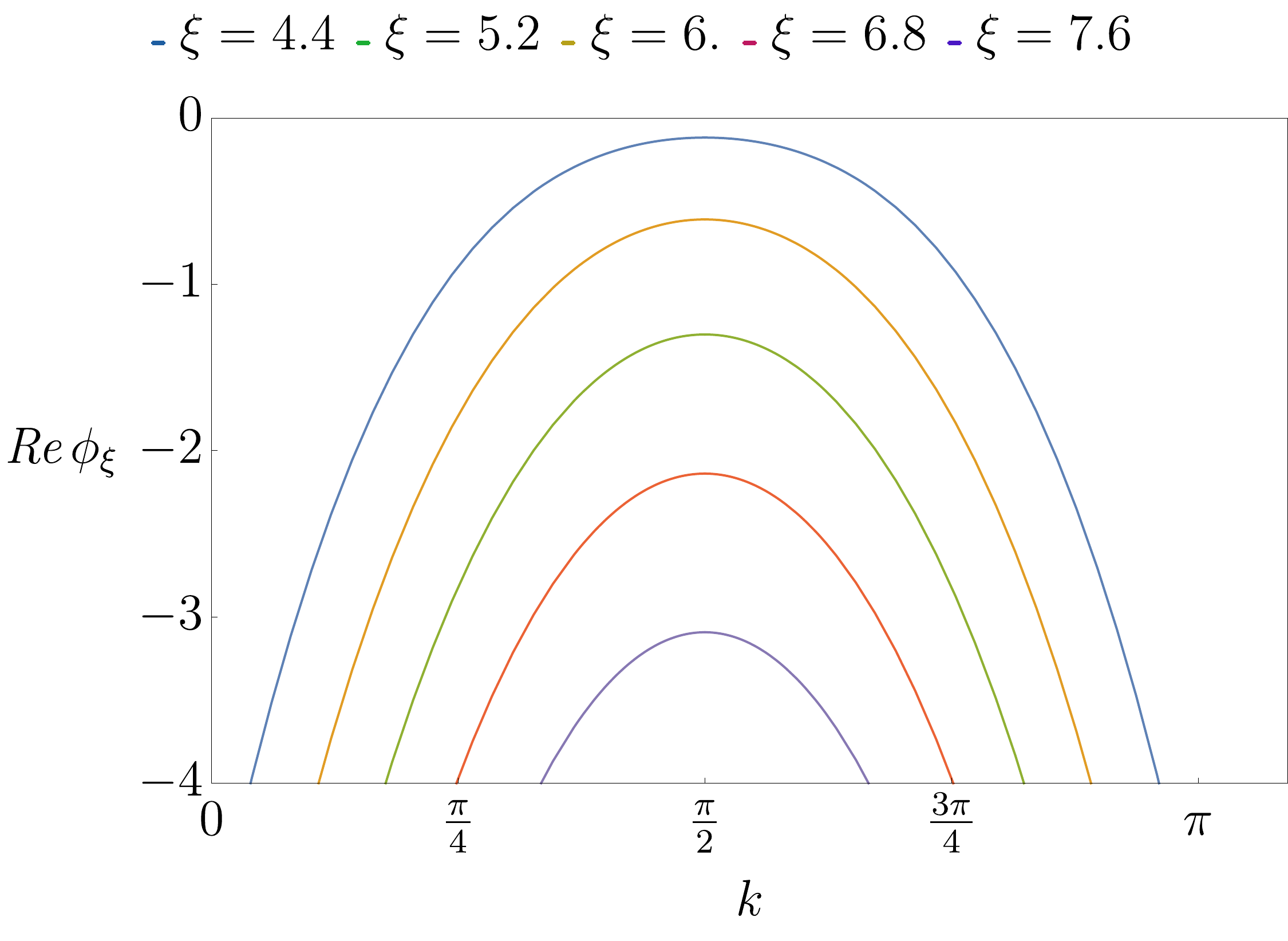}
	\end{center}
	\caption{(Left) Path used for the computation of the aymptotic of the integral \eqref{eq:shifted_w_space} for $\xi=4.2$ as defined by the Eqs.~\eqref{steepest} and \eqref{steepestconst}. The topology of the path neither depends on $h$ nor on $\xi$. (Right) $\real\phi_\xi$ for different $\xi$ evaluated along the parabolic like path passing through the saddle points (the black dot). Note how the maximum occurs at $k=\pi/2$ irrespective of $\xi>4$.}
	\label{fig:saddle_point}
\end{figure}

The steepest descent equation is
\beq\label{steepest}
	C_\xi - k\xi = 4\cos(k)\cosh(q)
\eeq
for some $k,q$-independent quantity $C_\xi$. The correct steepest descent curve is that which passes by $z=\pm\pi$ and by the closest saddle point, and where  $\real \phi_\xi$ is bounded. For $|\xi| > 4$, the closest saddle point of $\phi(z)$ is at $z_\xi = {\rm sgn}(\xi)\,z_{|\xi|}^*$ with $z^*_{|\xi|}=\pi/2-\ri\arccosh(|\xi|/4)$ (with $\arccosh(|\xi|/4)>0$). The choice of the sign of the imaginary part guarantees the fastest decay: the smallest maximum of $\real \phi_\xi$ on the steepest descent curve. The full curve $\gamma_\xi$ is then a concatenation of three curves $\gamma_{\xi,\pm}$, passing by $z=\pm\pi$, and $\gamma_\xi^*$, passing by the saddle point. The curves are determined by
\beq\label{steepestconst}
	C_{\xi,\pm} = \pm\xi \pi - 4,\qquad
	C_\xi^* = |\xi| \frc\pi 2.
\eeq
We show in Fig. \ref{fig:saddle_point} an example of the deformed contour  $\gamma_\xi = \gamma_{\xi,-}\cup\gamma_{\xi,+}\cup\gamma^*_{\xi}$ obtained as a union of steepest descent paths in the case $\xi>4$. 
The leading exponential decay depends on the analytic structure of $w(k)$ because where it has zeros the integrand has poles. As $t\rightarrow\infty$ we can write
\beq
\int_{\gamma_\xi}\frac{\ud z }{2\pi}\frac{\e^{t\phi_\xi(z)}}{1-\e^{w(z)}} \asymp \exp\left(\ri t\imag\phi_\xi(z^\epsilon_\xi)\right)\,\exp\left(t\real\phi_\xi(z^\epsilon_\xi) \right)+ \Lambda .
\eeq
The correction $\Lambda$ incorporates the residues of the singularities of $(1-\re^{w(k)})^{-1}$ swiped by the contour shift, which in general takes the form
\beq\label{Lambda}
\Lambda = \sum_{i:\,\re^{w(z_i)} = 1, \, z_i\in{\cal D}_\xi} K_i t^{n_i-1} \exp\big(t\phi_\xi(z_i)\big)
\eeq
for constants $K_i$ which can be evaluated explicitly (and which depend on $\xi$ for $n_i>1$), where $n_i\geq 1$ is the order of the zero $z_i$ of $w(z)$. The values of $z_i$ to be taken are those which lie within the swiped domain ${\cal D}_\xi$, for instance the domain bounded by the curved shown in Fig.~\ref{fig:saddle_point}, left. Using $\real\phi_\xi(z^*_\xi) = -|\xi| \arccosh(\xi/4)+4\sqrt{|\xi/4|^2-1}$ and $\imag\phi_\xi(z^*_\xi) = 2h - |\xi| \pi/2$ we obtain the estimate \eqref{exponentialbehaviour}.
In the main text we have explicitly analysed the thermal case but we can write down the general result valid on arbitrary homogenous GGEs as
\beqa\label{fullresulttransverse}
\lt|\bra\sigma_x^+(t)\sigma_0^-(0)\ket\rt|\asymp
\lt\{\begin{aligned}
	&\exp\lt( f_{x,t}[w]\rt) &&(|\xi|\leq 4)\\
	& \big(\exp\lt(-|x|\,M_\xi\rt)  + \Lambda\big)
	  \exp\lt(f_{x,t}[w]\rt)&&(|\xi|>4)
\end{aligned}
\rt.
\eeqa
where $\Lambda$ depends on the analytic structure of $w(k)$ and is defined in \eqref{Lambda}, $M_\xi$ is defined in \eqref{Mxi} and $f_{x,t}[w]$ in \eqref{eq:f_functional}.

\section{Numerical methods}
\label{app:numerical_methods}
\subsection{Pfaffian representation}
Consider the following quadratic fermionic Hamiltonian
\begin{equation}
	H = \sum_{x,y\in\Lambda}M_{x,y}a^\dag_x\,a_y + C
\end{equation}
In the thermodynamic limit, the bulk properties do not depend on the boundary conditions. The XX model eq. \eqref{eq:H_XX} is obtained with the choice $M_{x,y} = -2(\delta_{x,y-1} + \delta_{x-1,y} - \delta_{x,y})$ and $C = h N $. The task is to compute the following spin-spin correlation function
\begin{equation}
\rho_{+,-}(x,y;t)=\vev{\sigma^+_x(t) \sigma^-_y(0)}\label{eq:correlation_function}
\end{equation}
where the expectation value is taken on a GGE described by generalised temperatures $\ul{\beta}=(\beta_1,\dots\beta_N)$ corresponding to a set of extensive conserved quantities $\{Q_i\}_{i=1}^N$.
This correlation function is easily expressed in terms of the more fundamental correlators \cite{PhysRevA.2.1075, PhysRevA.3.786, PhysRevA.3.2137, PhysRevA.4.2331} 
\begin{equation}
	\rho^{i,j}(x,y;t) = \vev{\sigma^i_x(t)\sigma^j_y(0)}\quad.
\end{equation}
Indeed,
\begin{equation}
	\rho^{+,-}(x,y;t) = \frac{1}{4}\left[\vev{\sigma^1_x(t)\sigma^1_y} + \vev{\sigma^2_x(t)\sigma^2_y} - \ri\left(\vev{\sigma^1_x(t)\sigma^2_y}-\vev{\sigma^2_x(t)\sigma^1_y}\right)\right]\quad.
\end{equation}
As the Hamiltonian is invariant under rotations about the $3$-axis we can rotate spins $\sigma^2$ by $\pi/2$ and we obtain that the first to terms above are equal. The same rotation gives $\vev{\sigma^1_x(t)\sigma^2_y}= -\vev{\sigma^2_x(t)\sigma^1_y}$. Thus,
\begin{equation}
	\rho^{+,-}(x,y;t) = \frac{1}{2}\left[\rho^{1,1}(x,y;t)- \ri \rho^{1,2}(x,y;t)\right]
\end{equation}
and we only need two fundamental correlators. Introducing the following operators
\begin{subequations}
	\begin{equation}
		A_x = a^\dag_x + a_x
	\end{equation}
	\begin{equation}
		B_x = a^\dag_x - a_x
	\end{equation}
\end{subequations}
we rewrite the fundamental correlators as follows
\begin{subequations}
	\begin{equation}
		4\rho^{1,1}(x,y;t) = \vev{A_x(t) \prod_{0\leq z< x}A_z(t) B_z(t) \prod_{0\leq z'< y}A_{z'} B_{z'} A_y}
	\end{equation}
	\begin{equation}
		4\ri\rho^{1,2}(x,y;t) = \vev{A_x(t) \prod_{0\leq z< x}A_x(t) B_z(t) \prod_{0\leq z'< y}A_{z'} B_{z'} B_y}\quad.
	\end{equation}
\end{subequations}
Note that $\anticommuator{A_x}{B_y} = 0$ while $\anticommuator{A_x}{A_y} = -\anticommuator{B_x}{B_y} = 2\delta_{x,y}$ so that these operators are distinct fermions (not canonically normalised). Caianiello and Fubini show that the trace of such non-normalised fermions is a Pfaffian \cite{Caianiello1952}, for which there exist efficient numerical algorithms \cite{Wimmer_2012}. We recap how to arrive at Caianiello-Fubini formula below as we find it instructive and not enough advertised. Given an anti-symmetric matrix $A$ of size $2N\times 2N$ its determinant can be proven to be the square root of a polynomial in the matrix entries (see \cite{ledermann_1993} for an elementary proof). This polynomial is called Pfaffian and it is defined as \cite{muir1920theory}
\begin{equation}
	\pf M = \frac{1}{2^N N!}\sum_{\sigma\in\mathcal{S}_{2N}}\epsilon_\sigma\,M_{\sigma(1),\sigma(2)}\dots M_{\sigma(2N-1),\sigma(2N)}\quad .
\end{equation}
where $\mathcal{S}_{2N}$ is the the group of permutations of $2N$ elements and $\epsilon_\sigma$ is the sign of the permutation $\sigma$.
An equivalent way to write the Pfaffian is to notice that in the sum over all permutation there are redundant terms: indeed, we can permute the factors in $N!$ possible ways and due to the anti-symmetry of the matrix we can also permute indices giving rise to a $2^N$ factor (without affecting the overall sign). Taking into account this one can write
\begin{equation}
	\pf M = \sum_{\substack{i_1<j_1\dots i_{2N}<j_{2N}\\
		i_1<i_2<\dots<i_{2N}}}\epsilon_\sigma\, M_{i_1, j_1}\dots M_{i_{2N-1}, j_{2N}}\quad .\label{eq:pfaff2}
\end{equation}
where the sign of the permutation is still computed according to whether the sequence of indices is even or odd with respect to the standard order $1,\dots,2N$. The Pfaffian can be computed also in a recursive fashion. To show this let us define fermionic operators $\psi_x$ such that
\begin{equation}
	\anticommuator{\psi_x}{\psi_y}=2(xy)\in\mathbb{C}
\end{equation}
where the symbol $(xy)$ is simply a notation for the result of the anticommutation and can be any number. The problem is to compute the following trace
\begin{equation}
	\frac{1}{4}\Tr\{\psi_1,\dots \psi_{2N}\}\equiv (1\dots 2N)\quad.\label{eq:trace_pfaff}
\end{equation}
Using the anticommutation relations multiple times we obtain
\begin{equation}
	(1\dots 2N) = \sum_{k=2}^{2N}(-1)^k(1k)(2\,3\dots k-1\,k+1\dots 2N)
	\label{eq:pfaff_recursion}
\end{equation}
and by simple recursion
\begin{equation}
	(1\dots 2N) = \sum_{\substack{i_1<j_1\dots i_{2N}<j_{2N}\\
			i_1<i_2<\dots<i_{2N}}}\epsilon_\sigma(i_1 j_1)\dots(i_N j_N)\quad.\label{eq:corr_pfaff}
\end{equation}
Comparing \eqref{eq:corr_pfaff} and \eqref{eq:pfaff2} we see that if we define the anti-symmetric matrix with upper diagonal elements given by
\begin{equation}
	M_{i j} = (i j)
\end{equation}
the trace of products of fermionic operators becomes a Pfaffian. Further,  from eq. \eqref{eq:pfaff2} we see that
\begin{equation}
	\pf M = \sum_{k=2}^{2N}(-1)^k M_{ik}\pf M_{\hat{1}\hat{k}}\label{eq:pfaffian}
\end{equation}
where $M_{\hat{i}\hat{j}}$ is the matix $M$ with both rows and columns $i$ and $j$ removed.
This formula, when the trace in \eqref{eq:trace_pfaff} is substituted by a finite entropy state $\vev{\circ}_\rho$ represented by density matrices of the form \eqref{eq:free_GGE}, is still valid provided we use $\vev{ij}=\vev{\psi_i \psi_j}$ instead of $(ij)$ . In the XX model it is possible to compute exactly these fundamental correlaors between the $A$'s and the $B$'s. On GGEs as in eq. \eqref{eq:GGE} elementary steps give
\begin{subequations}
	\begin{equation}
		\vev{A_x(t)A_y}=-\vev{B_x(t)B(y)}=\sum_{k=0}^{N-1}g_{xk}C_kg_{ky}
	\end{equation}
	\begin{equation}
		\vev{A_x(t)B(y)}=-\vev{B_x(t)A(y)}=\sum_{k=0}^{N-1} g_{xk} S_k g_{ky}\quad .
	\end{equation}
\end{subequations}
The matrices above satisfy
\begin{subequations}
	\begin{equation}
		\sum_{x=0}^{N-1} g_{kx}A_{xy} = E_k g_{ky} \quad g^\dag = g^{-1}
	\end{equation}
while
	\begin{equation}
	C_k = \frac{\cosh\left( \ri E_k t - \frac{1}{2}W_k\right)}{\cosh\left(\frac{1}{2}W_k\right)}	
	\end{equation}
	\begin{equation}
	S_k = \frac{\cosh\left( -\ri E_k t + \frac{1}{2}W_k\right)}{\cosh\left(\frac{1}{2}W_k\right)}	\quad .
	\end{equation}
\end{subequations}
The function $E_k$ is the dispersion relation \eqref{eq:dispersion_relation} while $W_k$ is the function characterizing the GGE, the same in \eqref{eq:GGE} and \eqref{eq:free_GGE}.
Specialisation to finite temperature states is again given setting $W_k=\beta E_k$. For the actual numerical calculation of the Pfaffian in \eqref{eq:pfaffian} we have followed the methods of Ref. \cite{Wimmer_2012}.
\subsection{An alternative for time-independent correlators}
Define the $2N$ dimensional vectors of operators,
\begin{equation}
    \ul{V}=\begin{pmatrix}
    \ul{a}\\ \ul{a}^\dag
    \end{pmatrix},\quad\ul{V}^\dag=\begin{pmatrix}
    \ul{a}^\dag \\ \ul{a}
    \end{pmatrix} 
\end{equation}
These satisfy,
\begin{equation}
    \{V^\dag_z, V_{z'}\}=\delta_{z,z'}
\end{equation}
Now we consider the general correlator,
\begin{equation}
   G_{zz'} =  \Tr\left\{	\left(\prod_{i=1}^{n}\e^{M_i}\right)V^\dag_z\left(\prod_{i=1}^{m}\e^{M'_i}\right)V_{z'}	\right\}
\end{equation}
where $M, M'$ are $2N\times 2N$ fermionic bilinear forms.
Suppose that,
\begin{equation}
    \commutator{M_i}{\ul{V}}=\Omega_i \ul{V}, \quad \commutator{M'_i}{\ul{V}}=\Omega'_i \ul{V}
\end{equation}
In this way, 
\begin{equation}
    \e^{-M_i}V_z\e^{\mathcal{M}_i} = \left(\e^{\Omega_i}\ul{V}\right)_z, \quad \e^{-M'_i}V_z\e^{M'_i} = \left(\e^{\Omega'_i}\ul{V}\right)_z
\end{equation}
and,
\begin{equation}
    \commutator{M_i}{\ul{V}^\dag}=\sigma_i\Omega_i^*\ul{V}^\dag, \quad \commutator{M'_i}{\ul{V}^\dag}=\sigma_i\Omega'^*_i \ul{V}^\dag
\end{equation}
where $\sigma_i=-1$ if the $M_i$ is hermitian $\sigma_i=1$ if it is anti-hermitian.
Using this we can write (summation over repeated indices),
\begin{align}
    &\Tr\left\{	\left(\prod_{i=1}^{n}\e^{M_i}\right)V^\dag_z\left(\prod_{i=1}^{m}\e^{M'_i}\right)V_{z'}	\right\}
    \nonumber
    \\
    =&\Tr\left\{	\left(\prod_{i=1}^{n}\e^{M_i}\right)\left(\prod_{i=1}^{m}\e^{M'_i}\right)V_{z'}\left(\prod_{i=n}^{1}\e^{\sigma_i\Omega^*_i}\right)_{zl}V^\dag_{l}	\right\}
    \nonumber
    \\
    \nonumber
    =&\Tr\left\{	\left(\prod_{i=1}^{n}\e^{M_i}\right)\left(\prod_{i=1}^{m}\e^{M'_i}\right)\left(\prod_{i=n}^{1}\e^{\sigma_i\Omega^*_i}\right)_{zl}(\delta_{zl}-V^\dag_l V_{z'})	\right\}
    \nonumber
    \\
    =&-\Tr\left\{	\left(\prod_{i=1}^{n}\e^{M_i}\right)\left(\prod_{i=1}^{m}\e^{M'_i}\right)\left(\prod_{i=n}^{1}\e^{\sigma_i\Omega^*_i}\right)_{zl}V^\dag_l V_{z'}	\right\}\nonumber
    \\
    &+\Tr\left\{	\left(\prod_{i=1}^{n}\e^{M_i}\right)\left(\prod_{i=1}^{m}\e^{M'_i}\right)\left(\prod_{i=n}^{1}\e^{\sigma_i\Omega^*_i}\right)_{zz'}	\right\}
    \nonumber
    \\
    =&-\Tr\left\{	\left(\prod_{i=1}^{n}\e^{M_i}\right)V^\dag_s\left(\prod_{i=1}^{m}\e^{M'_i}\right)\left(\prod_{i=m}^{1}\e^{\sigma_i\Omega'^*_i}\right)_{ls}\left(\prod_{i=n}^{1}\e^{\sigma_i\Omega^*_i}\right)_{zl} V_{z'}	\right\}
    \nonumber
    \\
    &+\Tr\left\{	\left(\prod_{i=1}^{n}\e^{M_i}\right)\left(\prod_{i=1}^{m}\e^{M'_i}\right)\left(\prod_{i=n}^{1}\e^{\sigma_i\Omega^*_i}\right)_{zz'}	\right\}
    \end{align}
From this we get,
\begin{align}
    \left[\delta_{zs}+\left(\prod_{i=n}^{1}\e^{\sigma_i\Omega^*_i}\prod_{i=m}^{1}\e^{\sigma_i\Omega'^*_i}\right)_{zs}\right]G_{sz'}=\left(\prod_{i=n}^{1}\e^{\sigma_i\Omega^*_i}\right)_{zz'}\Tr\left\{	\left(\prod_{i=1}^{n}\e^{M_i}\right)\left(\prod_{i=1}^{m}\e^{M'_i}\right)	\right\}
\end{align}
and using Klich formula \cite{Klich2003, Klich_2014} we transform the trace into a determinant,
\begin{equation}
   G_{zz'}= \left[\prod_{i=m}^{1}\e^{\sigma_i\Omega'^*_i}+\prod_{i=1}^{n}\e^{-\sigma_i\Omega^*_i}\right]^{-1}_{zz'}
  \det\left(1+	\left(\prod_{i=1}^{n}\e^{M_i}\right)\left(\prod_{i=1}^{m}\e^{M'_i}\right)	\right)
  \label{eq:exact_two_point_function}
\end{equation}
which is a very simple, efficient and compact formula to study time-independent correlation functions. We have compared \eqref{eq:exact_two_point_function} with results from the pfaffian method when $t=0$.

\end{appendices}


\end{document}